\documentclass{article}
\usepackage{amsfonts}
\usepackage{amsmath}
\usepackage{color}
\usepackage{pdfpages}
\usepackage{subfig}
\usepackage{hyperref}
\usepackage{cite}

\newcommand*{\figuretitle}[1]{%
	{\centering
		\textbf{#1}
		\par\medskip}
}

\begin{document}

\begin{titlepage}
\centering

\title{Affine quantisation of the Brans--Dicke theory:\\
	Smooth bouncing and the equivalence between Einstein and Jordan frames}

\author{E. Frion\footnote{frion.emmanuel@hotmail.fr}\\
\textit{PPGCosmo, CCE, Universidade Federal do Esp\'irito Santo,}\\ \textit{Vit\'oria, 29075-910, Esp\'irito Santo, Brazil}\\
\textit{Institute of Cosmology and Gravitation, University of Portsmouth}\\
\textit{Portsmouth PO1 2EG, United Kingdom}
\vspace{0.5cm}\\
C.R. Almeida\footnote{cralmeida00@gmail.com}\\
\textit{COSMO - Centro Brasileiro de Pesquisas F\'isicas}\\
\textit{Rio de Janeiro, 22290-180, Brazil}\\
\textit{Max Planck Institute for the History of Science}\\
\textit{14195 Berlin, Germany}
\vspace{0.5cm}
}
\clearpage\maketitle
\thispagestyle{empty}

\begin{abstract}
	In this work, we present a complete analysis of the quantisation of the classical Brans-Dicke Theory using the method of affine quantisation in the Hamiltonian description of the theory. The affine quantisation method is based on the symmetry of the phase-space of the system, in this case the (positive) half-plane, which is identified with the affine group. We consider a Friedmann-Lema\^itre-Robertson-Walker spacetime, and since the scale factor is always positive, the affine method seems to be more suited than the canonical quantisation for our Quantum Cosmology. We find the wave function of the Brans-Dicke universe, and its energy spectrum. A smooth bounce is expected at the semi-classical level in the quantum phase-space portrait. We also address the problem of equivalence between the Jordan and Einstein frames.
\end{abstract}

\end{titlepage}

\section{Introduction}
\label{Introduction}

After the formulation of general relativity (GR), some modified theories arose in an attempt to explain open problems in cosmology, such as inflation and the observed accelerated expansion. One of the oldest modifications of GR is the Brans-Dicke theory (BDT), proposed in the early 1960s by Carl H. Brans and Robert H. Dicke \cite{Brans-Dicke}, in which there is a non-minimal time-dependent coupling of the long-range scalar field with geometry, that is, with gravity. The BDT also introduces an adimensional constant $\omega$ such that, for a constant gravitational coupling, GR is recovered at the limit  $\omega \rightarrow \infty$, if the trace of the energy-momentum tensor is not null \cite{Faraoni98,Faraoni99,Chauvineau}. Today it is well known that, classically, the BDT is practically indistinguishable from GR, with the constant $\omega$ estimated to be over $40,000$ \cite{Will,Avilez}. Interestingly, the Brans-Dicke scalar field arises naturally in superstring cosmology, associated with the dilaton, which couples directly with the matter field \cite{Wands}. The dilaton is equivalent to the graviton for a theory with dynamical gravitational coupling. In spite of the fact that the BDT is classically no different from GR, the quantum treatment can reveal new dynamics for the primordial Universe. There are also claims that the BDT can not reproduce GR for a scale-invariant matter content. In fact, in this case, it has been shown that $\omega$ can display various effects depending on its value, such as a symmetry breaking resulting in a binary phase structure. However, for a strong coupling $\omega \rightarrow \infty$, the BDT reproduces GR only in the quantised version \cite{pal2}.

With the assumption that quantum effects cannot be ignored at early stages of the Universe, the quantisation of the classical BDT in its Hamiltonian description is relevant to better undestand this era. We will assume a minisuperspace, a configuration with reduced degrees of freedom for homogeneous cosmologies, which can be understood as a projection of the whole superspace, containing only the largest wavelength modes of the size of the Universe \cite{Kerbrat}. Minisuperspaces are considered to be toy models, since they reduce the superspace, that is, the observable universe on largest scales, which have infinite degrees of freedom. However, it is still a fairly good approximation of the superspace to study certain properties. This allows one to target specific behaviors such as the dynamics of the volume of the Universe, or to investigate the nature of the initial singularity and the inflationary phase.

 We choose to explore the quantisation of the BDT with the affine quantisation instead of the canonical one, since the domain of the variables involved (scale factor and scalar field) is the real half-line, and its phase space can be identified with the affine group. With this, we also  avoid the operator-ordering problem arising in the case of canonical quantisation (see, for example, discussions in \cite{Isham,DeWitt}). The affine quantisation is also equipped with a ``de-quantisation" map that allows us to obtain classical expressions from quantum operators. In the canonical quantisation, the classical measurements are obtained by the expectation values of the classical observables, but in the affine quantisation the classical system is recovered with possible corrections through this de-quantisation map, called quantum corrections or lower symbols. While being a fairly recent subject of interest in cosmology, the affine quantisation points to interesting applications, such as the removal of divergences in non-renormalisable theories \cite{Klauder1,Klauder2} or the non-singular expanding (and possibly cyclic) universes \cite{Fanuel}.

This work, in which we will investigate the quantisation of the BDT applying the affine method, is a continuation of the analysis initiated in \cite{Carla tese}. We present the wave function for a Brans-Dicke universe, and we draw its quantum phase space. Then, it is shown that a bounce is expected, avoiding the initial singularity. We also raise the question about the equivalence between the Jordan and Einstein frames. This paper is organised as follows. In Section \ref{BDT}, we review the classical derivation of the BDT with a perfect fluid introduced via the Schutz formalism. In Section \ref{Aff. Quantisation}, we introduce the affine quantisation method as well as a more direct way to obtain classical estimates: the quantum phase-space portraits. In Section \ref{BDT quantisation}, we apply the affine quantisation to the BDT to obtain the Wheeler-DeWitt equation in the Jordan frame and in the Einstein frame. Finally, we derive the semi-classical version of the Hamiltonian constraint in both frames. In the last Section, we present our results and discuss the dependence of the parameters on the solutions.

\section{The Brans-Dicke theory with a perfect fluid}
\label{BDT}

The Brans-Dicke theory is characterised by the introduction of a scalar field non-minimally coupled to gravity, and it is described by the gravitational Lagrangian
\begin{equation}
\label{Lagrangian}
\mathcal{L}_{G} = \sqrt{-g}\left\{ \varphi R-\omega \frac{\varphi _{;\rho }\varphi^{;\rho }}{\varphi }\right\}\, .
\end{equation}
The Brans-Dicke coupling parameter $\omega$ is chosen to be a constant in this work. Let us consider a homogeneous and isotropic universe,
\begin{eqnarray}
\label{metric FLRW}
ds^{2}= N^{2} \left( t\right) dt^{2} - a^{2} \left( t\right) \left[dx^{2}+dy^{2}+dz^{2}\right] \, ,
\end{eqnarray}
where $N$ and $a$ are respectively the lapse function and the scale factor. Then, the Lagrangian (\ref{Lagrangian}) becomes
\begin{equation}
\label{Lagrangian 2}
\mathcal{L}_{G}=\frac{1}{N}\left\{ 6\left[ \varphi a\dot{a}^{2}+a^{2}\dot{a}%
\dot{\varphi}\right] -\omega a^{3}\frac{\dot{\varphi}^{2}}{\varphi }\right\} \,,
\end{equation}
where we have already discarded the surface terms. The Lagrangian of the system is completed with a matter component, which we will consider to be a radiative perfect fluid, defined by the equation of state $p=\rho/3$. 

Let us use the Schutz formalism to introduce the perfect fluid \cite{Schutz}, in which the four-velocity of a baryonic perfect fluid is described by four potentials,\footnote{There are six potentials in total, but they reduce to four in a homogeneous and isotropic medium.} the specific enthalpy $\mu$ and the entropy $s$ of the fluid and another two with no clear physical meaning, let us call them $\epsilon$ and $\theta$. After some considerations \cite{Alvarenga, Pedram}, the matter Lagrangian becomes
\begin{equation}
\label{matter Lagrangian}
\mathcal{L}_{M} = - \frac{1}{3} \left( \frac{3}{4} \right)^{4} \frac{a^{3}}{N^{3}} \left( \dot{\epsilon} + \theta \dot{s} \right)^{4} e^{- 3s} \,.
\end{equation}
Since we are interested in the quantum corrections of this system, we must describe the theory with the Hamiltonian formalism. To do so, let us write the Lagrangians above as functions of the conjugate momenta, defined by
\begin{eqnarray}
\label{momentum}
p_{q} = \frac{\partial \mathcal{L}}{\partial \dot{q}} \,.
\end{eqnarray} 
With this, from (\ref{matter Lagrangian}) we obtain the matter super-Hamiltonian \cite{nelson} \footnote{The Hamiltonian defined on the minisuperspace, where the space-like metric and non-gravitational fields belong to a finite set and their conjugate momentum is identically zero.}
\begin{equation}
\label{matter Hamiltonian 1}
\mathcal{H}_{M} = - p_{\epsilon}^{\frac{4}{3}} a^{-1} e^{s} \,, 
\end{equation} 
where $p_{\epsilon} = -N \rho_{0} U^{0} a^{3}$, with  $\rho_{0}$ the rest mass density of the fluid and $U$ the four-velocity. Let us introduce the following canonical transformations \cite{Rubakov}
\begin{equation}
\label{canonical transf.}
T= - p_{s} e^{-s} p_{\epsilon}^{-\frac{4}{3}} \quad; \quad p_{T} = p_{\epsilon}^{\frac{4}{3}} e^{s} \quad; \quad \overline{\epsilon} = \epsilon - \frac{4}{3} \frac{p_{s}}{p_{\epsilon}} \quad; \quad \overline{p}_{\epsilon} = p_{\epsilon} \,.
\end{equation}
Then, the super-Hamiltonian for the matter component becomes
\begin{equation}
\label{matter Hamiltonian 2}
\mathcal{H}_{M}=-\frac{N}{a} \,p _{T} \,.
\end{equation}%

The Hamiltonian for the gravitational part is given by the Legendre transformation of $\mathcal{L}_{G}$
\begin{equation}
\mathcal{H}_{G}= \dot{a} p_{a} + \dot{\varphi} p_{\varphi} -\mathcal{L}_{G} \,,
\end{equation}
where the conjugate momenta are
\begin{eqnarray}
	p_{a} = \frac{6}{N}\left(2\varphi a \dot{a} +a^2 \dot{\varphi}\right) \quad ; \quad 
	p_{\varphi } = \frac{6}{N}a^2 \dot{a} - 2\frac{\omega}{N}a^{3} \frac{\dot{\varphi}}{\varphi} \,.
\end{eqnarray}
Expressing the generalised velocities in terms of the momenta, we obtain
\begin{eqnarray}
	\dot{a} = \frac{\omega N}{(3+2\omega)\varphi a} \left(\frac{p_{a}}{6}+\frac{\varphi p_{\varphi }}{2\omega a}\right) \quad ; \quad \dot{\varphi} = \frac{N \varphi}{2(3+2\omega ) a^3} \left(\frac{a\,p_{a}}{\varphi}-2p_{\varphi}\right) \,,
\end{eqnarray}
which, after some algebra, gives us
\begin{equation}
	\mathcal{H}_{G} = \frac{N}{3+2\omega} \left[ \frac{\omega 
	}{12\varphi a}p _{a}^{2}+\frac{1}{2a^{2}}p _{a}\,p _{\varphi }-\frac{%
		\varphi }{2a^{3}} p_{\varphi }^{2}\right] \,.
\end{equation}
Therefore, the Hamiltonian of the BDT is given by
\begin{equation}
\label{Total Hamillt.}
\mathcal{H}=N\left\{ \frac{1}{\left( 3+2\omega \right) }\left[ \frac{\omega }{12\varphi a} p_{a}^{2}+\frac{1}{2a^{2}} p_{a}\,p _{\varphi }-\frac{\varphi }{2a^{3}} p_{\varphi }^{2}\right] -\frac{1}{a} p_{T} \right \} \,,
\end{equation}%
where $p_T$, $p_{a}$ and $p_{\varphi }$ are the conjugate
momenta associated with the matter component, the scale factor $a$ and the field $\varphi $, respectively. 

The classical Hamiltonian constraint $\mathcal{H} \approx 0$ still holds (notice that here $\approx$ means ``weakly equal", so that $\mathcal{H}$ is a first class constraint, i.e. its Poisson brackets with other constraints are vanishing on the constrained space) for the BDT with a perfect fluid \cite{Dirac,Barrow}. Thus, we have
\begin{equation}
\label{Hamiltnian constraint}
\frac{\omega}{12 \varphi} p_{a}^{2} + \frac{1}{2a} p_{a} p_{\varphi} - \frac{\varphi}{2a^{2}} p_{\varphi}^{2}  = (3+ 2\omega) p_{T} \,.
\end{equation}
The quantisation of this constraint results in the Wheeler-DeWitt equation. We can interpret it as a Schr\"odinger-like equation and, from it, obtain the cosmological scenarios at a quantum level \cite{Carla2}. Now, instead of the canonical quantisation used in \cite{Carla2}, we will introduce another quantisation method, based on the symmetry group of the system's phase space. This kind of quantisation is completed with a quantum phase space portrait, which accounts for a quantum correction to the classical trajectories of the theory, that we will use to analyse the BDT at early cosmological times.

\section{The affine quantisation}
\label{Aff. Quantisation}

\subsection{Mathematical background}
\label{Mathematical backg.}

First, let us introduce the affine quantisation method mentioned earlier. The model requires the scale factor and the scalar field, our two dynamical variables, to be positive, with the zero value being a geometrical singularity. Thus, the phase space is a four-dimensional space which is the cartesian product of two half-planes,\footnote{In the case of radiative matter, at least \cite{Carla2}.} 
\begin{equation}
\Pi^{2}_{+}:=\left\{(a,p_{a}) \times (\varphi, p_{\varphi}) \,|\, a>0, \varphi >0 \,,\, p_{a}, p_{\varphi} \in \mathbb{R} \right\} \,.
\end{equation}
Since it is a cartesian product, we can analyse each half-plane separately. Thus, we will present the theory behind this method of quantisation for a generic phase space, and then apply it to our specific case (for a more detailed presentation, see \textit{e.g.} \cite{Bergeron5,Bergeron6}). 

The half-plane $\Pi_{+}:=\{(q,p)\,|\, q>0\,,\,p\in \mathbb{R} \}$ with a multiplication operation defined by
\begin{equation}
\label{multiplication} 
\left(q,p \right) \left(q_{0},p_{0} \right) = \left( qq_{0}, \frac{p_{0}}{q}+p \right) \,; \quad q \in \mathbb{R}_{+}^{*} \,,\quad p \in \mathbb{R} \, ,
\end{equation}
is identified with the affine group Aff$_{+}(\mathbb{R})$ of the  real line. The group acts on $\mathbb{R}$ as follows
\begin{equation}
\label{group action}
(q,p) \cdot x = \frac{x}{q} + p \quad,\quad  \forall \, x \in \mathbb{R} \,. 
\end{equation}
On a physical level, one can interpret (\ref{group action}) as a contraction/dilation (depending on if $q>1$ or $q<1$) of space plus a translation.
We shall equip the half-plane with the measure $dq \, dp$, which is invariant under the left action of the affine group on itself \cite{Carla3}. 

Rigorously, the affine quantisation is a covariant integral method, that combines the properties of symmetry from the affine group with all the resources of integral calculus. This method makes use of \textit{coherent states} \cite{Gazeau1} to construct the quantisation map, whose definition is connected with the symmetry of the phase-space, as we will see. First, let us explain the integral quantisation method. Given a group $G$ and a unitary irreducible representation (UIR) of it, the quantisation map transforms a classical function (or distribution) into an operator using a bounded square-integrable operator $M$ and a measure $d\nu$, such as
\begin{equation}
\label{identity resolution}
\int_{G} M(g) \,d\nu(g) = I \,,
\end{equation}
where $g \in G$, $M(g) = U(g) M U^{-1}(g)$. This is the resolution of the identity for the operator $M$. With this, from a classical observable $f(g)$, we obtain the corresponding operator
\begin{equation}
\label{int. quant. map}
A_{f} = \int_{G} M(g) \, f(g) \, d\nu(g) \,.
\end{equation}

For the affine group, that is $G= \text{Aff}_{+}(\mathbb{R})$, we have two non-equivalent UIR $U_{\pm}$, plus a trivial one $U_{0}$ \cite{Aslaksen,Isham84}. Let us choose $U = U_{+}$, which acts on the Hilbert space 
$L^{2}(R_{+}^{\ast},dx/x^{\alpha +1})$ as 
\begin{equation}
\label{U+}
(U(q,p) \,\psi)(x)=\frac{e^{i px}}{\sqrt{q^{-\alpha}}}\psi \left( \frac{x}{q} \right)\,.
\end{equation}
We choose the operator $M$ such as
\begin{equation}
\label{fiducial vectors}
M = | \psi \rangle \langle \psi | \quad; \quad \psi \in L^{2} \left(R_{+}^{\ast},\frac{dx}{x^{\alpha +1}} \right) \cap L^{2} \left(R_{+}^{*},\frac{dx}{x^{\alpha +2}} \right) \,.
\end{equation}
The normalised vectors $\psi$ are arbitrarily chosen providing the square-integrability condition (\ref{fiducial vectors}), and they are known as \textit{fiducial vectors}. For simplicity, we will consider only real fiducial vectors and will choose $\alpha=-1$. The action (\ref{U+}) of the UIR of $U$ over fiducial vectors produces the quantum states
\begin{equation}
\label{ACS definition}
|q,p\rangle := U(q,p)|\psi\rangle \,.
\end{equation}
These states are called \textit{affine coherent states} (ACS) or \textit{wavelets}. It is easy to show that 
\begin{equation}
\label{affine res. identity}
\int_{\Pi_{+}}|q,p\rangle\langle q,p|\,\frac{dq \, dp}{2\pi c_{-1}}=I\,,
\end{equation}
where the constant $c_{-1}$ depends on the choice of $\psi$, and is defined as
\begin{equation}
\label{c_gamma}
c_{\gamma} = c_{\gamma}(\psi) := \int_{0}^{\infty} |\psi(x)|^{2} \, \frac{dx}{x^{2+\gamma}} \,.
\end{equation}
Hence, the quantisation maps (\ref{int. quant. map}) becomes
\begin{equation}
\label{aff. quant. map} 
f(q,p)\ \mapsto\ A_{f}=\int_{\Pi_{+}}f(q,p) \,|q,p\rangle\langle
q,p| \, \frac{dq dp}{2\pi c_{-1}}\,.
\end{equation}
With this, one can easily verify that the quantisation of the elementary functions position $q^{\beta}$ (for any $\beta$), momentum $p$ and kinetic energy\footnote{up to a factor.} $p^{2}$ yields
\begin{equation}
\label{quantum operators} 
A_{q^{\beta}} = \frac{c_{\beta-1}}{c_{-1}} \, \hat{Q}^{\beta}\quad;\quad A_{p}=-i\frac{\partial}{\partial x}= \hat{P} \quad ; \quad
A_{p^{2}}= \hat{P}^{2} + \frac{c^{(1)}_{-3}}{c_{-1}} \hat{Q}^{-2} \,,
\end{equation}
with $\hat{Q}$ being the position operator defined by $\hat{Q}f(x)=xf(x)$ and the constant $c^{(1)}_{-3}$ is defined as
\begin{equation}
\label{c^beta_gamma}
c^{(\beta)}_{\gamma} (\psi) := \int_{0}^{\infty} \Big|\psi^{(\beta)}(x) \Big|^{2} \, \frac{dx}{x^{2+\gamma}} \,.
\end{equation}

Notice that, in this affine quantisation method, the only dependence on the fiducial vector $\psi$ is in the constant coefficients of the quantum operators. Thus, the arbitrariness of $\psi$ does not play a fundamental role in the quantisation. This is an advantage to be explored. For example, we can adjust the fiducial vectors to regain the self-adjoint character of the operator $p^{2}$ \cite{Carla3} . Choosing $\psi$ such that $ 4c^{(1)}_{-3} \geq 3c_{-1}$, the kinetic operator becomes essentially self-adjoint \cite{Reed2}, which is a desired characteristic since an Hermitian operator must be self-adjoint. An Hermitian operator can be obtained by imposing boundary conditions. However, there is a continuous infinity of possible boundaries, thus the choice of a representation is arbitrary (this is the \textit{operator ordering} problem of the canonical quantisation). In the affine quantisation, the choice of a fiducial vector can naturally result in an essentially self-adjoint operator, which means there is only one possible extension of it and, therefore, no need to impose boundary conditions. We stress, however, that choosing fiducial vectors is not the same as choosing boundary conditions. Self-adjointness is a well-known problem in the canonical quantisation of this theory, and it has been studied extensively in \cite{Carla2}. However, with the affine quantisation we naturally recover the quantum symmetrisation of the classical product momentum position 
\begin{equation}
\label{operator qp} 
qp \quad \mapsto \quad A_{qp} = \frac{c_0}{c_{-1}}\frac{\hat{Q} \hat{P} + \hat{P} \hat{Q}}{2} \,,
\end{equation}
up to a constant that once again depends on the choice of the fiducial vector.

\subsection{Quantum phase-space portraits}
\label{semclassport}

The construction of the affine quantisation method presented in the previous section using coherent states allows us to define a ``de-quantisation" map, named quantum phase-space portrait, in a very obvious way: by calculating the expectation value of an operator with respect to the coherent states. That is, given a quantum operator $A_{f}$, we obtain a classical function $\check{f}$ such that 
\begin{equation}
\label{checkf}
\check{f} (q,p) = \langle q,p | \, A_{f} \, | q,p\rangle \,.
\end{equation}
If the operator is obtained from a classical function $f$, as suggested in the notation, then $\check{f}$ is a quantum correction or lower symbol of the original $f$ \cite{Lieb}. It corresponds to the average value of $f(q,p)$ with respect to the probability density distribution  
\begin{equation}
\label{prob. distr.} 
\rho_{\phi}(q,p)=\frac{1}{2\pi c_{-1}} | \langle q,p|\phi \rangle |^{2} \,,
\end{equation}
with $|\phi \rangle = |q^{\prime},p^{\prime} \rangle $. We can also define the time evolution of the distribution (\ref{prob. distr.}) with respect to time through a Hamiltonian operator $\hat{H} = A_{H}$, using the time evolution operator $e^{-i \hat H t}$. Then,
\begin{equation}
\label{time prob. distr.} 
\rho_{\phi}(q,p,t):=\frac{1}{2\pi c_{-1}}|\langle q,p|e^{-i \hat H t}|\phi\rangle|^{2} \,.
\end{equation}
Thus, if you consider the operator $M=\rho$, the lower symbol of $A_f$ becomes \cite{Bergeron5}
\begin{equation}
	\check{f}(z)=\int tr\left(\rho(z) \rho(z')\right) f(z')\frac{d^2 z^{'}}{\pi} \, ,
\end{equation}
with $tr$ the trace. From the resolution of the identity \ref{identity resolution}, one finds $tr\left(\rho(z) \rho(z')\right)$ is a probability distribution of the phase-space, and $\check{f}$ is indeed an average measurement of the classical $f$.

From equation (\ref{checkf}), using (\ref{aff. quant. map}), the quantum correction $\check{f}$ of a classical function $f$ is then
\begin{eqnarray}
\nonumber
\check{f}(q,p) = \frac{1}{2\pi c_{-1}} \int_{-\infty}^{\infty} \int_{0}^{\infty} \frac{dq^{\prime} \, dp^{\prime}}{qq^{\prime}} \int_{0}^{\infty} \int_{0}^{\infty} dx \, dx^{\prime} f(q^{\prime}, p^{\prime}) \left[ e^{ip(x^{\prime} - x)}  \right.
\\
\label{checkf formulae}
\left. \times \, e^{-ip^{\prime}(x^{\prime} - x)} \psi \left(\frac{x}{q}\right) \psi \left(\frac{x^{\prime}}{q}\right) \psi \left(\frac{x}{q^{\prime}}\right) \psi \left(\frac{x^{\prime}}{q^{\prime}} \right) \right] \,.
\end{eqnarray}
Thus, it is not necessary to find the operator $A_{f}$ of a classical function $f$ to obtain its lower symbol. One can use the above formula (\ref{checkf formulae}) to do so. For example, the quantum correction of the classical functions $q^{\beta}$, $p$ and $p^{2}$ are given by
\begin{equation}
\label{semi-class. correc.} 
\check{q^{\beta}} = \frac{c_{\beta-1}c_{-\beta-2}}{c_{-1}} \, q^{\beta} \quad; \quad  \check{p}= p \quad; \quad \check{p^2}= p^2 + \left(c_{-2}^{(1)} + \frac{c_{0} c^{(1)}_{-3}}{c_{-1}}\right) \frac{1}{q^2}\,,
\end{equation}
with the constants $c_{\gamma}$ and $c^{(\beta)}_{\gamma}$ defined in (\ref{c_gamma}) and (\ref{c^beta_gamma}), respectively. Notice that the corrections also depend on the choice of specific fiducial vectors to determine these constants.

\section{The affine quantisation of the BDT}
\label{BDT quantisation}

\subsection{Quantisation in the Jordan Frame}

Now that we have introduced the affine quantisation method and the quantum phase-space portrait coming from it, we can apply the method to the BDT presented in Section \ref{BDT}, since the variables $a$ and $\varphi$ are both positively defined. However, the Schutz variable associated to the fluid has the whole real line as its domain and therefore we cannot apply the affine method in it. Nevertheless, we can use another integral quantisation method based on the Weyl-Heisenberg group, which acts on the real line \cite{Gazeau2}. Here we could also use the canonical quantisation for this variable, since it works just fine for parameters in the whole line, a domain that does not have any singularity and, therefore, no problems of self-adjointness.\footnote{Using the Weyl-Heisenberg method can give us the advantage of introducing another constant that depends on the fiducial vector chosen in the quantisation. This can be an asset used to adjust energy levels, for example.} In both cases, we have
\begin{equation}
p_{T}  \quad \mapsto \quad \hat{P}_{T} = -i \, \frac{\partial}{\partial T} \quad; \quad p_{T} \quad \mapsto \quad \check{p}_{T} = p_{T} = E \,.
\end{equation}

To build the coherent states of the variables $a$ and $\varphi$, let us name the respective fiducial vectors as $\psi_{a}$ and $\psi_{\varphi}$, which are \textit{a priori} not the same. Then, the coherent states are given by
\begin{eqnarray}
\label{coherent state a}
|a, p_{a} \rangle = U_{a} \, |\psi_{a} \rangle  \quad &\Rightarrow& \quad \langle x \, |a, p_{a} \rangle = \frac{e^{ip_{a} x}}{\sqrt{a}} \, \psi_{a} \left( \frac{x}{a} \right)
\\
\label{coherent state varphi}
|\varphi, p_{\varphi} \rangle = U_{\varphi} \, |\psi_{\varphi} \rangle  \quad &\Rightarrow& \quad \langle y \, |\varphi, p_{\varphi} \rangle  = \frac{e^{ip_{\varphi} y}}{\sqrt{\varphi}} \, \psi_{\varphi} \left( \frac{y}{\varphi} \right) \,.
\end{eqnarray}
With this, the quantisation of equation (\ref{Hamiltnian constraint}) results in the following Wheeler-DeWitt equation:
\begin{eqnarray}
\nonumber
\left\{- \omega \lambda_{1} \frac{1}{\varphi} \partial_{a}^{2} + \left(\omega \lambda_{2} - \lambda_{3} \right) \frac{1}{\varphi a^{2}} - \lambda_{4} \frac{1}{a} \partial_{a} \partial_{\varphi} + \lambda_{5} \frac{\varphi}{a^{2}} \partial_{\varphi}^{2} + \quad \quad \quad \quad \right.
\\
\label{WDW equation}
\left.  + \lambda_{6} \frac{1}{a^{2}} \partial_{\varphi} \right\} \Psi(a, \varphi, T) = -i \left(3 + 2\omega\right) \partial_{T} \Psi (a, \varphi, T) \,,
\end{eqnarray}
where $\Psi (a,\varphi,T)$ is the wave function. The constants $\lambda_{i}$ are given by
\begin{eqnarray}
\nonumber
\lambda_{1} = \frac{1}{12 c_{-1}(\varphi)} \quad&;& \quad \lambda_{2} = \frac{1}{12 c_{-1}(\varphi)} \frac{c_{-3}^{(1)}(a)}{c_{-1}(a)} \,;
\\
\label{lambdas}
\lambda_{3} = \frac{1}{2} \frac{c_{-3}(a)}{c_{-1}(a)} \frac{c_{-2}^{(1)}(\varphi)}{c_{-1}(\varphi)} \quad&;& \quad \lambda_{4} = \frac{1}{2c_{-1}(a)}\,;
\\
\nonumber
\lambda_{5} = \frac{1}{2} \frac{c_{-3}(a)}{c_{-1}(a)} \frac{c_{0}(\varphi)}{c_{-1}(\varphi)} \quad &;& \quad \lambda_{6} = \frac{1}{2} \frac{c_{-3}(a)}{c_{-1}(a)} \frac{c_{0}(\varphi)}{c_{-1}(\varphi)} + \frac{1}{4c_{-1}(a)} \,,
\end{eqnarray}
and we defined
\begin{equation}
c_{\gamma}^{(j)}(a) = \int_{0}^{\infty} [\psi_{a}^{(j)}(x)]^{2} \frac{dx}{x^{2+ \gamma}} \quad; \quad c_{\gamma}^{(j)}(\varphi) = \int_{0}^{\infty} [\psi_{\varphi}^{(j)}(x)]^{2} \frac{dx}{x^{2+ \gamma}} \,.
\end{equation}

If we choose $\psi_{a} = \psi_{\varphi}$, then $c_{\gamma}^{(j)}(a) = c_{\gamma}^{(j)}(\varphi) = c_{\gamma}^{(j)}$. So, let us choose a fiducial vector such that
\begin{equation}
\psi_{a} = \psi_{\varphi} = \frac{9}{\sqrt{6}} \, x^{\frac{3}{2}} \, e^{-\frac{3x}{2}} \,.
\end{equation}
With these vectors, we have $c_{-2} = c_{-1} = 1$, and $c_{-3}^{(1)} = 3/4$, which, as mentioned before, is a necessary condition for the quantised kinetic energy to be an essentially self-adjoint operator \cite{Reed2}. In turn, this gives the us the Wheeler-DeWitt equation in the Jordan frame
\begin{eqnarray}
	\label{WDW equation Jordan}
	\left\{- \frac{\omega}{12} \frac{1}{\varphi} \partial_{a}^{2} + \left( \frac{\omega}{16} - \frac{3}{4} \right) \frac{1}{\varphi a^{2}} - \frac{1}{2a} \partial_{a} \partial_{\varphi} + \frac{\varphi}{a^{2}} \partial_{\varphi}^{2} + \frac{5}{4a^{2}} \partial_{\varphi} \right\} \Psi = -i \left(3 + 2\omega\right) \partial_{T} \Psi \,.
\end{eqnarray}
From this equation, absorbing the constant $12\left(3 + 2\omega\right) \omega^{-1}$ into the temporal parameter, that is, accounting it as energy, we find the Hamiltonian for the BDT in the Jordan frame to be
\begin{equation}
\label{Hamiltonian Jordan}
H_{J} = \frac{1}{\varphi} \partial_{a}^{2} - \frac{12}{\omega} \left( \frac{\omega}{16} - \frac{3}{4} \right) \frac{1}{\varphi a^{2}} + \frac{6}{\omega a} \partial_{a} \partial_{\varphi} - \frac{12}{\omega} \frac{\varphi}{a^{2}} \partial_{\varphi}^{2} - \frac{15}{\omega a^{2}} \partial_{\varphi} \,.
\end{equation}
It is easy to see that the Hamiltonian (\ref{Hamiltonian Jordan}) is essentially self-adjoint for the usual measure $da \,d\varphi$ on the Hilbert space, as expected. One can notice that equation (\ref{WDW equation Jordan}) is not separable. We can work around this problem by considering the Einstein frame instead.

\subsection{Conformal transformation of affine operators}
\label{Conf. transf.}

The Jordan and Einstein frames are related to each other by a conformal transformation given by $g_{\mu \nu} = \phi^{-1} \, \tilde{g}_{\mu \nu}$, where $g_{\mu \nu}$ and $\tilde{g}_{\mu \nu}$ represent the metric tensors in each frame, respectively. Thus, before analysing the equivalence between these frames, let us first comment on how affine operators change with a conformal transformation. 

As opposed to what happens in the canonical quantisation (see \cite{Carla tese}), the affine operators are uniquely defined by equation (\ref{aff. quant. map}). Also, if $A_{f}$ is the operator obtained from a classical function $f(q,p)$, with $q$ being a positive-defined variable and $p$ its associated momentum, then for a general conformal scaling factor  $\Omega(q)$ on the domain, we have
\begin{equation}
\label{fAg neq Afg}
\Omega^{2}(q) A_{f} \neq A_{\Omega^{2}(q) f} \,.
\end{equation}
Therefore, we need to be careful when we quantise models related by conformal transformations. Even if the constraint obtained from an Hamiltonian is classical, we cannot cancel overall coefficients (for instance, the factor $1/b$ in equation \ref{Hamilt. const. Einstein frame}).To illustrate this, let us give an example. Consider $\Omega^{2}(q) = q$ and $f(q,p) = p$. The operator $A_{\Omega^{2} f}$ is given by (\ref{operator qp}), and then
\begin{equation}
A_{\Omega^{2} f} = A_{qp} = \frac{c_{0}}{c_{-1}} \, \frac{\hat{Q}\hat{P} + \hat{P}\hat{Q}}{2} \neq \hat{Q}\hat{P} = qA_{p} = \Omega^{2}(q) A_{f}\,.
\end{equation}

This means that classically, it is always possible to cancel non-null coefficients,
however, quantising the constraint in different frames can result in very different scenarios, because of (\ref{fAg neq Afg}). 
In conclusion, we cannot cancel out non-null functions before quantising to compare the quantisation of two different frames connected by a transformation of coordinates.

\subsection{Quantisation in the Einstein frame}
\label{Einstein frame}

Since the seminal paper of Brans and Dicke \cite{Brans-Dicke}, we know that two formulations of the theory (and in fact, for every scalar-tensor theory) are possible. These formulations, related by a conformal transformation, are the target of a long debate on which of these frames is physically relevant. Some authors claim they are equivalent classically but should be different at the quantum level \cite{Artymowski,Banerjee}, while others claim that both are equivalent at classical and quantum level \cite{Carla4,Kamenschchik,Pandey,ohta}. Some also claim that the equivalence is broken by off-shell one-loop quantum corrections, but recovered on-shell\cite{ruf}. Since theoretical predictions depend entirely on the conformal frame we are working on, a natural question arising is if there is a preferred frame or not, and which one is the most suitable to observations. In the Jordan frame, we found the differential equation governing the wave function evolution (\ref{WDW equation Jordan}), however as a crossed term appeared in the partial derivatives, finding a solution can be  difficult. Let us now analyse the problem in the Einstein frame instead.

The Brans-Dicke Lagrangian, with a non-minimally coupled scalar field, is given by (\ref{Lagrangian}), and by using the conformal transformation, $g_{\mu \nu} = \varphi^{-1} \, \tilde{g}_{\mu \nu}$, where $g_{\mu \nu}$ is the metric in the non-minimal coupling frame, the Lagrangian reads as
\begin{equation}
\label{GR Lagrangian}
\mathcal{L}_{G} = \sqrt{-\tilde{g}} \left[\tilde{R} - \biggr(\omega + \frac{3}{2}\biggl) \frac{\varphi_{;\rho} \, \varphi^{;\rho}}{\varphi^{2}}\right] \,,
\end{equation}
which is the Lagrangian for General Relativity with a minimally coupled scalar field. The Lagrangian (\ref{Lagrangian}) is written in the Jordan frame, and (\ref{GR Lagrangian}) is written in the Einstein frame. The conformal transformation is given by the change of coordinates
\begin{equation}
\label{conformal trans.}
N^{\prime} = \varphi^{\frac{1}{2}} N \quad; \quad b = \varphi^{\frac{1}{2}} a \quad ; \quad \varphi^{\prime} = \varphi \,,
\end{equation}
and, applying these to (\ref{Lagrangian 2}), we obtain
\begin{equation}
\mathcal{L}_{G} = \frac{1}{N^{\prime}} \left[ 6b \dot{b}^{2} - \left( \omega + \frac{3}{2}\right)b^{3} \left(\frac{\dot{\varphi^{\prime}}}{\varphi^{\prime}} \right)^{2} \right] \,.
\end{equation}
The total Hamiltonian is thus
\begin{equation}
\label{Hamiltonian Einstein}
H_{T} = N^{\prime} \left( \frac{p_{b}^{2}}{24b} - \frac{\varphi^{\prime \, 2}}{2(3+ 2\omega)b^{3}} p_{\varphi^{\prime}}^{2} - \frac{p_{T}}{b} \right) \,,
\end{equation}
and the constraint $H_{T} =0$ gives us\footnote{We keep the 1/b factor in order to avoid inconsistences in the quantisation (see the discussion in Subsection \ref{Conf. transf.}).}
\begin{equation}
\label{Hamilt. const. Einstein frame}
\frac{p_{b}^{2}}{24b} - \frac{\varphi^{\prime \, 2}}{2(3+ 2\omega)b^{3}} p_{\varphi^{\prime}}^{2} = \frac{p_{T}}{b} \,.
\end{equation}

In order to quantise equation (\ref{Hamilt. const.  Einstein frame}), it is necessary to know the Hilbert space in the Einstein frame. From the change of variables (\ref{canonical transf.}), the measure becomes
\begin{equation}
da \, d\varphi = \varphi^{\prime \,-\frac{1}{2}} db \, d\varphi^{\prime} \,.
\end{equation}
Thus, the Hilbert space for the coordinates $(b,\varphi^{\prime})$ is $L^{2} (\mathbb{R}_{+}^{*} \times \mathbb{R}_{+}^{*},\varphi^{\prime \,-\frac{1}{2}} dbd\varphi^{\prime})$. Then, according to definition (\ref{fiducial vectors}), the fiducial vectors $\psi_{\varphi^{\prime}}$ are defined on another Hilbert space:
\begin{equation}
\psi_{\varphi^{\prime}} \in L^{2} \left(R_{+}^{\ast},\frac{dx}{x^{\frac{1}{2}}} \right) \cap L^{2} \left(R_{+}^{*},\frac{dx}{x^{\frac{3}{2}}} \right) \,.
\end{equation}
With this measure, the operator associated with the kinetic energy, is given by
\begin{equation}
A_{p^{2}} = -\partial_{\varphi^{\prime}}^{2} + \frac{1}{2\varphi^{\prime}} \partial_{\varphi^{\prime}} + \left( \frac{c_{-5/2}^{(1)}(\varphi^{\prime})}{c_{-1/2}(\varphi^{\prime})} - \frac{3}{8} \right) \frac{1}{\varphi^{\prime \,2}} \,,
\end{equation}
which is already self-adjoint. 

Now, for the coordinate $b$, using (\ref{aff. quant. map}), we obtain
\begin{eqnarray}
A_{b^{-1}p_{b}^{2}} &=& - \frac{1}{c_{-1}(b)} \frac{1}{b} \partial_{b}^{2} + \frac{1}{c_{-1}(b)} \frac{1}{b^{2}} \partial_{b} - \left(\frac{1-c_{-4}^{(1)}(b)}{c_{-1}(b)} \right) \frac{1}{b^{3}} \,.
\end{eqnarray}
For the coordinate $\varphi^{\prime}$, we get
\begin{eqnarray}
A_{\varphi^{\prime \, 2} p_{\varphi^{\prime}}^{2}}  &=& -\frac{11}{8}\frac{c_{3/2}}{c_{-1/2}} +\frac{c_{-1/2}^{(1)}}{c_{-1/2}}-\frac{3}{2}\frac{c_{3/2}}{c_{-1/2}} \varphi^{\prime} \partial_{\varphi^{\prime}} -\frac{c_{3/2}}{c_{-1/2}}\varphi^{\prime 2} \partial_{\varphi^{\prime}}^2 \,. 
\end{eqnarray}
Then, the quantisation of equation (\ref{Hamilt. const. Einstein frame}) results in
\begin{eqnarray}
\label{WDW equation Einstein gen.}
\left\{- \varpi \partial_{b}^{2} + \frac{\varpi}{b} \partial_{b} + \left( \tilde{\lambda}_{1} \varpi  + \tilde{\lambda}_{2} \right) \frac{1}{b^{2}} + \frac{\tilde{\lambda}_{3}}{b^{2}} \left(\varphi^{\prime \, 2} \partial_{\varphi^{\prime}}^{2} 
 + \frac{3}{2} \varphi^{\prime} \, \partial_{\varphi^{\prime}} \right) \right\}\Psi = -24\varpi i \,\partial_{T}\Psi \,, 
\end{eqnarray}
with $\varpi = \omega + \frac{3}{2}$, and $\tilde{\lambda}_{i}$ are given by
\begin{eqnarray}
\nonumber
\tilde{\lambda}_{1} &=& c_{-4}^{(1)}(b) -1  \,;
\\
\tilde{\lambda}_{2} &=& \frac{3}{4} \frac{ c_{-4}(b)}{ c_{-1/2}(\varphi^{\prime})} \left( \frac{11}{8}c_{3/2}(\varphi^{\prime}) - c_{-1/2}^{(1)}(\varphi^{\prime}) \right) \,;
\\
\nonumber
\tilde{\lambda}_{3} &=& \frac{c_{-4}(b) \, c_{3/2}(\varphi^{\prime})}{c_{-1/2}(\varphi^{\prime})} \,.
\end{eqnarray}

On the other hand, one can change variables as in (\ref{conformal trans.}) directly on (\ref{WDW equation}). This yields
\begin{eqnarray}
\nonumber
\label{WDW equation Einstein}
\left[- \left(\omega \lambda_{1} + \frac{\lambda_{4}}{2} - \frac{\lambda_{5}}{4} \right) \partial_{b}^2 +\frac{\lambda_{5}-\lambda_{4}}{4}  \frac{1}{b} \partial_{b} + \left( \omega \lambda_{2} - \lambda_{3} \right) \frac{1}{b^2} \,+  \right.
\\
\left.  +\, \left(\frac{\lambda_{5}}{2}-\lambda_{4} \right) \frac{\varphi^{\prime}}{b}\partial_{b} \partial_{\varphi^{\prime}} + \lambda_{5} \frac{\varphi^{\prime 2}}{b^2} \partial^{2}_{\varphi^{\prime}} + \lambda_{6} \frac{\varphi^{\prime}}{b^2} \partial_{\varphi^{\prime}}  \right] \Psi = -i(3+2\omega)\partial_{T} \Psi \,,
\end{eqnarray}
with $\lambda_{i}$ given in (\ref{lambdas}). Notice that the coefficients $\lambda_{i}$ are in terms of $c_{\lambda}^{(i)}(a)$ and $c_{\lambda}^{(i)}(\varphi)$, while the coefficients in equation (\ref{WDW equation Einstein gen.}) are in terms of $c_{\lambda}^{(i)}(b)$ and $c_{\lambda}^{(i)}(\varphi^{\prime})$. Considering the freedom in the choice of the fiducial vectors,\footnote{The quantization is not determined by this choice, although there is an inequality constraint ($4c_{-3}^{(1)} \geq 3c_{-1}$) in order to obtain a Hermintian operator (see discussion at the end of section \ref{Mathematical backg.}).} and comparing equations (\ref{WDW equation Einstein gen.}) and (\ref{WDW equation Einstein}), we conclude that there is equivalence between Einstein and Jordan frames only if
\begin{align}
	\frac{\lambda_{5}}{2}-\lambda_{4} = 0 \quad \Rightarrow \quad c_{-3}(a) = 2 \frac{c_{-1}(\varphi)}{c_{0}(\varphi)} \,.
\end{align}
In a way, this result is similar to the one found in \cite{Carla3}, where it is concluded that the equivalence depends on the choice of \textit{ordering factors} for the canonical quantisation, which are related to the coefficients of the Hamiltonian operator. In our case, the unitary equivalence is then obtained if we impose some constraints on the fiducial vectors:
\begin{equation}
4\,c_{-3}^{(1)} \geq 3\,c_{-1} \, \quad \text{for} \quad \psi_{a}, \psi_{b}, \psi_{\varphi} \,; \quad \text{and} \quad c_{-3}(a) = \frac{2c_{-1}(\varphi)}{c_{0}(\varphi)} \,.
\end{equation}

Let us solve, without loss of generality, equation (\ref{WDW equation Einstein gen.}). We suppose the following separation of variables: $\Psi (b,\varphi ,t) := X(b)\,Y (\varphi)\, P(T)$. We obtain, for the function of time
\begin{equation}
P(T)=A \exp \left[{i \frac{ET}{24}} \right] \,,
\end{equation}
where $E/24$ is the energy constant. This results in the following system of partial differential equations:
\begin{eqnarray}
\label{Bessel equation}
\left\{-\partial_{b}^{2} + \frac{1}{b} \partial_{b} + \frac{1}{\varpi} \left[ \tilde{\lambda_{1}} \varpi + \tilde{\lambda_{2}} - \tilde{\lambda_{3}} k^{2} \right] \frac{1}{b^{2}}\right\} X(b) = E \, X(b) \,; \nonumber 
\\
\label{Euler equation}
\left\{\varphi^{2} \partial_{\varphi}^{2} + \frac{3}{2} \varphi \, \partial_{\varphi} \right\} Y(\varphi) = -k^{2} \, Y(\varphi) \,, 
\end{eqnarray}
with $k^{2}$ being a separation constant. The general solutions are given by
\begin{eqnarray}
\label{X_b}
X(b) &=& C_{1}\, b\, J_{\nu} \left(\sqrt{E}\,b\right)	+ C_{2} \, b\, Y_{\nu} \left(\sqrt{E} \, b\right) \,,
\\
\label{Y_varphi}
Y(\varphi) &=& D_{1} \, \varphi^{-\frac{1}{4}(\sqrt{1-16k^2}+1)} + D_{2} \, \varphi^{\frac{1}{4}(\sqrt{1-16k^2}-1)} \,,
\end{eqnarray} 
with $J_{\nu}$ and $Y_{\nu}$ the Bessel functions of first and second kind, respectively, $C_{1,2}$, $D_{1,2}$ are integration constants,
\begin{equation}
\label{Bessel index}
\nu = \sqrt{\frac{\left(1+\tilde{\lambda_{1}}\right) \varpi + \left(\tilde{\lambda_{2}} - \tilde{\lambda_{3}} k^{2}\right)}{\varpi}} \,,
\end{equation}
and $k^{2}<1/16$.
The wave-function of the Universe $\Psi_{T}(b, \varphi) = X(b)Y(\varphi)$ must be square-integrable. This is the reason for the choice of the limit set for the separation constant. Equation (\ref{Euler equation}) is known as Euler equation and the solution (\ref{Y_varphi}) corresponds to said limit of $k^{2}$. The solution for $k^{2} = 1/16$ gives similar results, however $k^{2}>1/16$ results in a non square-integrable wave-function. This is also the reason why we choose a negative sign for the separation constant. Also, since $Y_{n}$ blows up at the origin, we must take $C_{2}=0$. Now, let us consider the following transformation for the variable $\varphi$:
\begin{equation}
\sigma =  \ln \varphi \quad \Rightarrow \quad d\sigma = \frac{1}{\varphi} d\varphi \,.
\end{equation}
With this, the solution (\ref{Y_varphi}) becomes\footnote{With this, it becomes more evident why it is only square-integrable for $k^{2} < 1/16$.}
\begin{equation}
Y(\sigma) = D_{1} \, e^{-\frac{\sigma}{4} \left(\sqrt{1-16k^2}+1\right)} + D_{2} \, e^{\frac{\sigma}{4} \left(\sqrt{1-16k^2}-1\right)} \,.
\end{equation}
For the sake of simplicity, let us consider $D_{2} =0$. We construct the wave packet as 
\begin{equation}
\Psi = N \int_{-\frac{1}{4}}^{\frac{1}{4}} dk b\, J_{\nu} \left(\sqrt{E}\,b\right) e^{-\frac{\sigma}{4} \left(\sqrt{1-16k^2}+1\right)} \, e^{i\frac{E}{24}T} \,,
\end{equation}
where $N$ is a normalisation constant. Therefore, the norm of the wave packets is 
\begin{eqnarray}
\nonumber
\langle \Psi | \Psi \rangle = N^{2} \int\limits_{0}^{b_{0}} \int\limits_{0}^{\infty} \varphi^{-\frac{1}{2}}\, db \, d\varphi   \, \int\limits_{-\frac{1}{4}}^{\frac{1}{4}} \int\limits_{-\frac{1}{4}}^{\frac{1}{4}}  dk \,dk^{\prime} \, b^{2}\, e^{-\frac{\sigma}{2}}  \times
\\
\times\, J_{\nu} \left(\sqrt{E}\,b\right) J_{\nu^{\prime}} \left(\sqrt{E}\,b\right) e^{i \left(\frac{1}{4}\sqrt{|1-16k^{\prime \,2}}|- \frac{1}{4}\sqrt{|1-16k^2|} \right) \sigma} \,,
\end{eqnarray}
or, writing only in terms of $\sigma$,
\begin{eqnarray}
\nonumber
\langle \Psi | \Psi \rangle = N^{2} \int\limits_{0}^{b_{0}} \int\limits_{-\infty}^{\infty} db \,d\sigma \, \int\limits_{-\frac{1}{4}}^{\frac{1}{4}} \int\limits_{-\frac{1}{4}}^{\frac{1}{4}}  dk \,dk^{\prime} \, b^{2} J_{\nu} \left(\sqrt{E}\,b\right)  \times
\\
\times \, J_{\nu^{\prime}} \left(\sqrt{E}\,b\right) e^{i \left(\frac{1}{4}\sqrt{|1-16k^{\prime \,2}}|- \frac{1}{4}\sqrt{|1-16k^2|} \right) \sigma} \,,
\end{eqnarray}
where the prime on the $\nu$ indicates $\nu(k^{\prime})$ and we can take $b_{0}=1$ as the value of the scale factor today. Performing the integrals over $\sigma$ and $k^{\prime}$ gives
\begin{equation}
\label{norm}
\langle \Psi | \Psi \rangle = 8\pi N^{2} \int_{0}^{b_{0}} \int_{-\frac{1}{4}}^{\frac{1}{4}} db dk \, b^{2} \, J_{\nu} \left(\sqrt{E}\,b\right) J_{\nu} \left(\sqrt{E}\,b\right) \,.
\end{equation}
Now, we shall consider an approximation for the limit $\omega \gg k^2$. This approximation is relevant due to our understanding of today's estimate of the Brans-Dicke constant $\omega$. Notice that, in this limit, the Bessel index (\ref{Bessel index}) becomes $\nu = \sqrt{1 + \tilde{\lambda}_{1}}$ and then (\ref{norm}) becomes
\begin{equation}
\langle \Psi | \Psi \rangle = 8\pi N^{2} \int_{0}^{b_{0}} db \, b^{2} J_{\nu = \sqrt{1 + \tilde{\lambda}_{1}}} \left(\sqrt{E}\,b\right) J_{\nu = \sqrt{1 + \tilde{\lambda}_{1}}} \left(\sqrt{E}\,b\right) \,.
\end{equation}
The solution is given in terms of the regularised generalised hypergeometric function $ _2\tilde{F}_3$ \cite{nist} as
\begin{align}
\nonumber
\label{wave packet norm}
\langle \Psi | \Psi \rangle =& 4\sqrt{\pi } N^2 b_0^3 \,\, \Gamma \left(\nu+\frac{1}{2}\right) \Gamma \left(\nu+\frac{3}{2}\right) \left(b_0 \sqrt{E}\right)^{2 \nu} \times 
\nonumber 
\\ 
& _2\tilde{F}_3\left(\nu + \frac{1}{2},\nu +\frac{3}{2}; \nu+1, \nu +\frac{5}{2},2 \nu+1;-b_0^2 E\right) \,.
\end{align}
The regularised generalised hypergeometric functions are defined as the power series
\begin{equation}
_p\tilde{F}_q\left(a_{1}, ..., a_{p}; b_{1},..., b_{q}; z\right) := \frac{1}{\Gamma{(b_1)}...\Gamma{(b_q)}} \sum_{n=0}^{\infty} \frac{(a_1)_n ...(a_p)_n}{(b_1)_n ...(b_q)_n} \frac{z^n}{n!}\, ,
\end{equation}
with the recurrence relations 
\begin{equation}
(a_{j})_{0} = 1 \,; \quad \text{and} \quad (a_{j})_{n} = a_{j} \left(a_{j}+1\right) \left(a_{j}+2\right) ... \left(a_{j}+n-1\right) \,, \quad \text{for} \quad n\geq 1 \,.
\end{equation}
The norm of the wave packet becomes
\begin{align}
\label{wave packet norm 2}
	\langle \Psi | \Psi \rangle =& A \left(b_0 \sqrt{E}\right)^{2 \nu} \sum_{n=0}^{\infty} \frac{\left(\nu +\frac{1}{2}\right)_n }{\left(\nu +1\right)_n \left(2\nu +1\right)_n} \frac{(-b_0^2 E)^n}{n!} \, , 
\end{align}
where
\begin{align}
	A = 4\sqrt{\pi} N^{2} \frac{b_0^3}{\left(\nu +1\right)} \frac{\Gamma \left(\nu+\frac{1}{2}\right)}{\Gamma \left(\nu +2\right)\Gamma \left(2\nu+1\right)} \,.
\end{align}
Then, equation \eqref{wave packet norm 2} suggests that the energy spectrum is discrete. This means we can write $\langle \Psi | \Psi \rangle = \sum_{n} \langle \Psi_{n} | \Psi_{n} \rangle$, and the energy levels satisfy the equations
\begin{equation}
 \langle \Psi_{0} | \Psi_{0} \rangle = A  \left(b_0 \sqrt{E}\right)^{2 \nu}  \,,
\end{equation}
and, for a general $n \geq 1$,
\begin{equation}
\langle \Psi_{n} | \Psi_{n} \rangle = A  \left(b_0 \sqrt{E}\right)^{2 \nu} \sum_{n=0}^{\infty} \frac{\left(\nu+\frac{1}{2}\right)_n }{\left(\nu +1\right)_n \left(2\nu+1\right)_n} \frac{(-b_0^2 E)^n}{n!}  \,.
\end{equation}

\subsection{Quantum phase-space portrait of the BDT}
\label{Quantum phase-space BDT}

Let us consider the formalism introduced in section \ref{semclassport}. The constraint (\ref{Hamiltnian constraint}), $\mathcal{H}_{T} = 0$, can be rewritten in its semi-classical version using (\ref{checkf formulae}) to calculate each term. For the sake of simplicity, we will keep the same letter for the energy constant, so  $\check{p_{T}} = E$, and hence
\begin{eqnarray}
\label{energy constratint semiclas.}
\frac{\omega}{12} \, \frac{1}{\varphi} p_{a}^{2} + \left( \omega \kappa_{1} - \kappa_{2} \right) \frac{1}{a^{2} \varphi} + \frac{1}{2a} p_{a} p_{\varphi} - \kappa_{3} \frac{\varphi}{a^{2}} p_{\varphi}^{2} = (3+2\omega) E \,,
\end{eqnarray}
with the constants $\kappa_{i}$ being
\begin{eqnarray}
\nonumber
\kappa_{1} &=& \frac{1}{12} \left(\frac{c_{0}(a)c_{-3}^{(1)}(a)}{c_{-1}(a)} + c_{-2}^{(1)}(a) \right) \,;
\\
\nonumber
\kappa_{2} &=& \frac{1}{2} \, \frac{c_{0}(a) c_{-3}(a)}{c_{-1}(a)} \left(\frac{c_{0}(\varphi)c_{-3}^{(1)}(\varphi)}{c_{-1}(\varphi)} + c_{-2}^{(1)} (\varphi) \right) \,;
\\
\nonumber
\kappa_{3} &=& \frac{1}{2} \, \frac{c_{0}(a) c_{-3}(a)}{c_{-1}(a)} \frac{c_{0}(\varphi) c_{-3}(\varphi)}{c_{-1}(\varphi)} \,,
\end{eqnarray}
where $c_{\gamma}^{(j)}(a)$ and $c_{\gamma}^{(j)}(\varphi)$ are 
\begin{equation}
\label{c_gamma a and varphi}
c_{\gamma}^{(j)}(a) = \int_{0}^{\infty} [\psi_{a}^{(j)}(x)]^{2} \frac{dx}{x^{2+ \gamma}} \quad; \quad c_{\gamma}^{(j)}(\varphi) = \int_{0}^{\infty} [\psi_{\varphi}^{(j)}(x)]^{2} \frac{dx}{x^{2+ \gamma}} \,.
\end{equation}
If we choose $\psi_{a} = \psi_{\varphi}$, then $c_{\gamma}^{(j)}(a) = c_{\gamma}^{(j)}(\varphi) = c_{\gamma}^{(j)}$. With this in mind, let us choose a fiducial vector such that
\begin{equation}
\psi_{a} = \psi_{\varphi} = \frac{9}{\sqrt{6}} \, x^{\frac{3}{2}} \, e^{-\frac{3x}{2}} \,.
\end{equation}
With these vectors, we have $c_{-2} = c_{-1} = 1$, and $c_{-3}^{(1)} = 3/4$, the latter being a necessary condition for the quantised Hamiltonian to be an essentially self-adjoint operator \cite{Reed2}. We want this condition to hold even if we are not doing the quantisation explicitly, since the semi-classical trajectories are probabilistic along the path that a quantum state evolves. Then, (\ref{energy constratint semiclas.}) becomes
\begin{eqnarray}
\label{semiclassical constraint}
\frac{\omega}{12} \, \frac{1}{\varphi} p_{a}^{2} + \frac{9}{8} \left( \omega - 2 \right) \frac{1}{a^{2} \varphi} + \frac{1}{2a} p_{a} p_{\varphi} - 2 \frac{\varphi}{a^{2}} p_{\varphi}^{2} = (3+2\omega) E \,.
\end{eqnarray}
The expression (\ref{semiclassical constraint}) allows us to analyse the expected behaviour of the scale factor $a$ for the early univese, for a given initial value of the scalar field $\varphi (t_{0}) = \varphi_{0}$ and its momentum at this instant $p_{\varphi} (t_{0}) = p_{\varphi \, 0}$.  

Notice that equation (\ref{Hamiltnian constraint}) is the classical Hamiltonian constraint in the Jordan frame. To compare the expected behaviour of the scale factor in the Jordan frame with that in the Einstein frame, let us calculate the quantum phase-space portrait of equation (\ref{Hamilt. const. Einstein frame}), the Hamiltonian constraint in the Einstein frame. We have
\begin{eqnarray}
\left(b^{-1} p_{b}^{2} \right)\check{} &=& \frac{p_{b}^2}{b}  +\frac{c^{(1)}_{-1}(b)+c_1(b) \, c^{(1)}_{-4}(b)-c_1(b)}{c_{-1}(b)} \frac{1}{b^3} \,;
\\
\left(\varphi^{\prime 2} p_{\varphi^{\prime}}^{2}\right)\check{} &=& \frac{c_{3/2}(\varphi^{\prime}) \, c_{-7/2}(\varphi^{\prime})} {c_{-1/2}(\varphi^{\prime})}\,  \varphi^{\prime \, 2} \, p_{\varphi^{\prime}}^{2}(\varphi^{\prime}) + \left( \frac{c_{3/2}(\varphi^{\prime}) \, c_{-7/2}^{(1)} (\varphi^{\prime})} {c_{-1/2}(\varphi^{\prime})} + \right.
\nonumber 
\\
& &\left. + \frac{c_{-3/2}(\varphi^{\prime}) \, c_{-1/2}^{(1)}(\varphi^{\prime})}{c_{-1/2}(\varphi^{\prime})} - \frac{11}{8}\frac{ c_{3/2}(\varphi^{\prime}) c_{-3/2}(\varphi^{\prime})} {c_{-1/2}(\varphi^{\prime})} \right) \,.
\end{eqnarray}
Then, the quantum correction of (\ref{Hamilt. const. Einstein frame}) becomes
\begin{eqnarray}
\frac{3+2\omega}{24} \left[ p_{b}^2+\kappa_{4} \frac{1}{b^2} \right]  - \frac{1}{b^2}\left[\kappa_{5}\varphi^{\prime 2} p_{\varphi^{\prime}}^2  + \kappa_{6} \right] = (3+2\omega)E^{'},
\end{eqnarray}
with $E^{'}$ the energy, and the constants
\begin{eqnarray}
\nonumber
\kappa_{4} &=& \frac{c^{(1)}_{-1}(b)+c_1(b) \, c^{(1)}_{-4}(b) - c_1(b)}{c_{-1}(b)} \,; 
\\
\nonumber
\kappa_{5} &=& \frac{1}{2}\frac{c_{-4}(b) \, c_{1}(b)}{c_{-1}(b)} \, \frac{c_{3/2}(\varphi^{\prime}) \, c_{-7/2}(\varphi^{\prime})} {c_{-1/2}(\varphi^{\prime})} \,;
\\
\nonumber
\kappa_{6} &=& \frac{1}{2} \frac{c_{-4}(b) \,c_{1}(b)}{c_{-1}(b)} \, \frac{c_{3/2}(\varphi^{\prime})} {c_{-1/2}(\varphi^{\prime})} \, \left( c_{-7/2}^{(1)} (\varphi^{\prime}) + \frac{c_{-3/2}(\varphi^{\prime}) \, c_{-1/2}^{(1)}(\varphi^{\prime})}{c_{3/2}(\varphi^{\prime})}  - \frac{11}{8} c_{-3/2}(\varphi^{\prime}) \right) \,.
\end{eqnarray}
By choosing the fiducial vectors as before, we find
\begin{eqnarray}
	\label{semiclassical constraint einstein}
	\frac{3+2\omega}{24} p_{b}^2 + \left[\frac{1296-1500\sqrt{3\pi}+864\omega}{64}\right]\frac{1}{b^2}-\frac{525\sqrt{3\pi}}{16b^2}\varphi^{\prime 2} p_{\varphi^{\prime}}^2  = (3+2\omega)E^{'}.
\end{eqnarray}
Equations (\ref{semiclassical constraint}) and (\ref{semiclassical constraint einstein}) are the quantum corrections of the classical Brans-Dicke Theory described in the Jordan and Einstein frames, respectively. To understand the consequences of these corrections, let us build the quantum phase-space of the BDT in both these frames.

\section{Phase-space portraits}
\label{Results}

As mentioned before, in this section we present the quantum phase-space portraits coming from equations \eqref{semiclassical constraint} and \eqref{semiclassical constraint einstein}. The aim is to understand the behaviour of the scale factor $a$, which is connected to the volume of the Universe, so the phase-spaces shown here are with reference to this variable. Notice, however, that there are still other free parameters: the scalar field $\varphi$, the energy $E$ and the Brans-Dicke constant $\omega$. These parameters will be varied for the sake of understanding their influence on the issue. Without loss of generality, let us consider the initial state of the scalar field to be $\varphi_{0} =1$. 

\subsection{Jordan frame}

\begin{figure*}[htp!]
	\centering
	\figuretitle{Phase-space portrait for $p_{a}$ and $a$, varying the range of $p_{\varphi}$.}
	\subfloat{\includegraphics[width=0.5\textwidth]{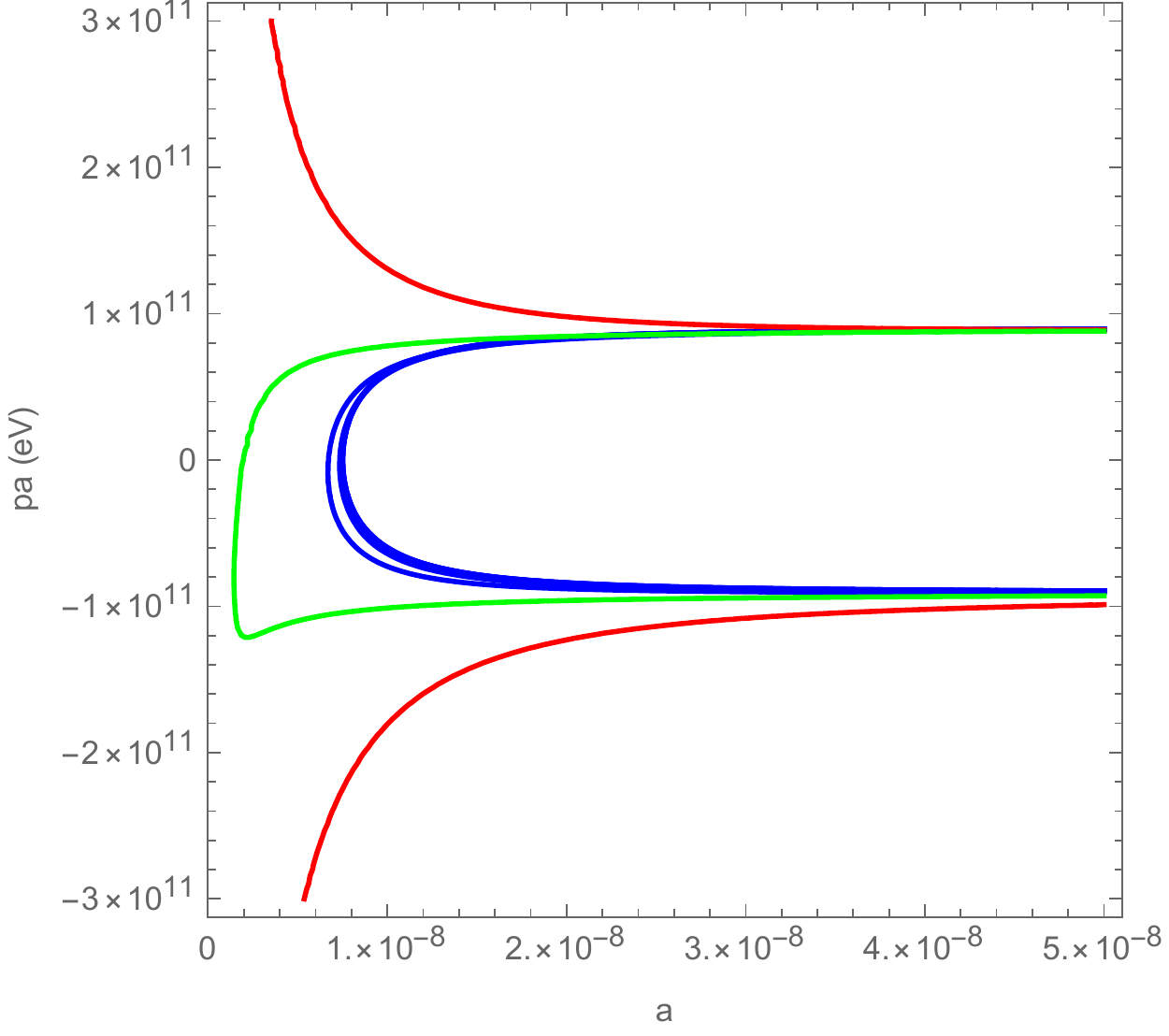}}
	\subfloat{\includegraphics[width=0.5\textwidth]{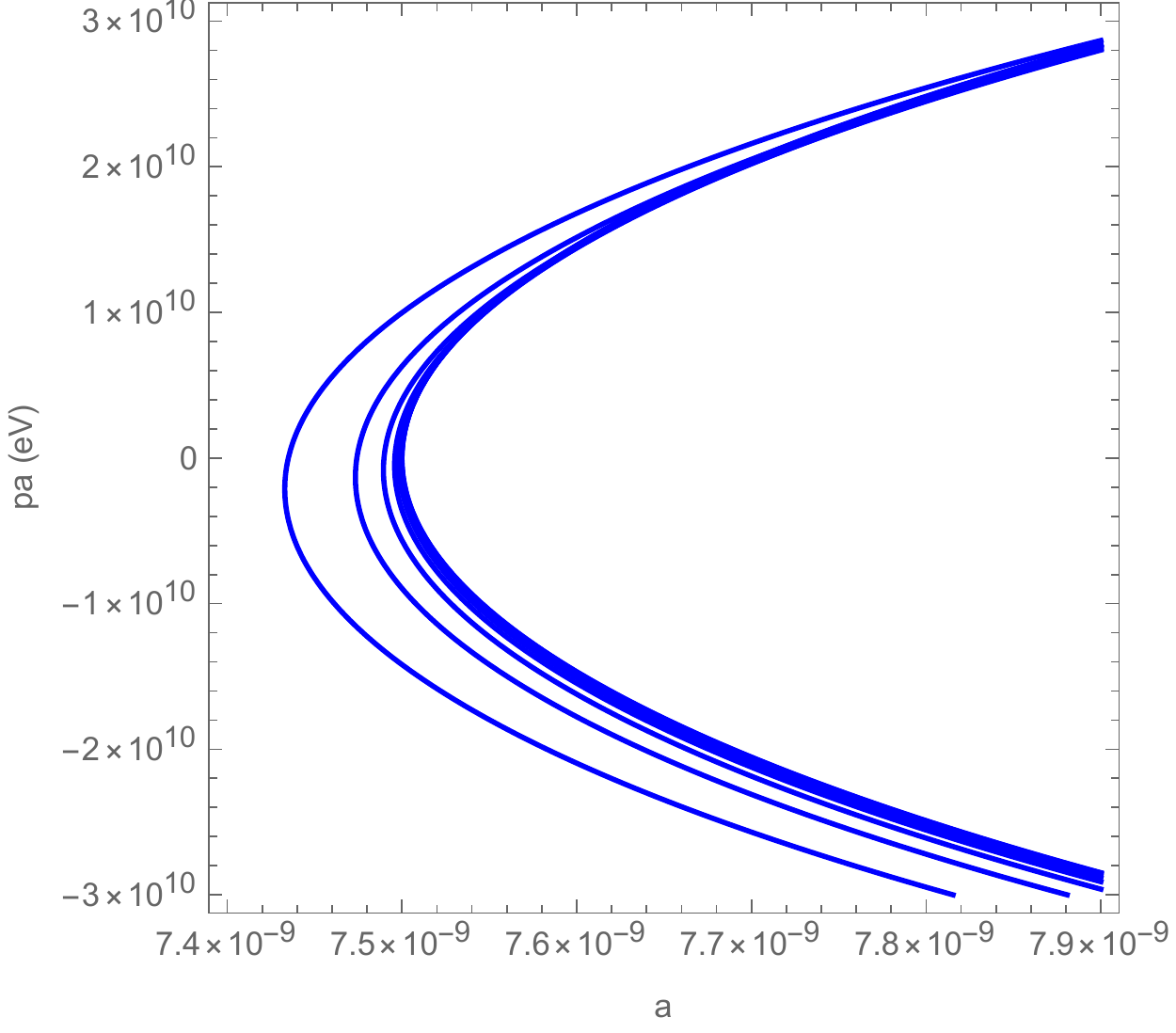}}\hfill
	\caption{Quantum phase-space of the scalar field in the Jordan frame, using $\omega=410,000$ and $E_{0}=10^{16}$. The left figure is for a range $1  \leq p_{\varphi} \leq 10^3$, while for the right figure the  range is smaller $1 \leq p_{\varphi} \leq 10^2$.}
	\label{Jordan moment. varying}
\end{figure*}

\textbf{Figure \ref{Jordan moment. varying}:} 

For the Jordan frame, let us set the energy at $E_{0}$ and construct the phase-space for a range of values of $p_{\varphi}$. Each curve represents a value for the velocity (momentum) of the scalar field. In each plot, we have a total of ten curves. For each curve, the less the minimum of the scale factor is, the higher $p_{\varphi}$ is. Notice that, up until an upper value for $p_{\varphi}$, the curves are of a smooth bouncing for the Universe, including solutions with a possible inflationary phase. Above a certain value of $p_\varphi$, divergent curves appear. If one assumes that this type of divergence does not describe a physical reality (favoring smoothness), then the scalar field must have a limit in momentum. Otherwise, this model predicts a singularity formed by an accelerated contraction of a prior universe, reaching null volume as the (modulus of the) momentum goes to infinity, followed by a decelerated inflation.\footnote{Notice that, we are reading the graphics in the clockwise direction.}

\begin{figure*}[htp!]
	\centering
	\figuretitle{Phase-space portrait for $p_{a}$ and $a$, varying the value of $\omega$.}
	\subfloat{\includegraphics[width=0.5\textwidth]{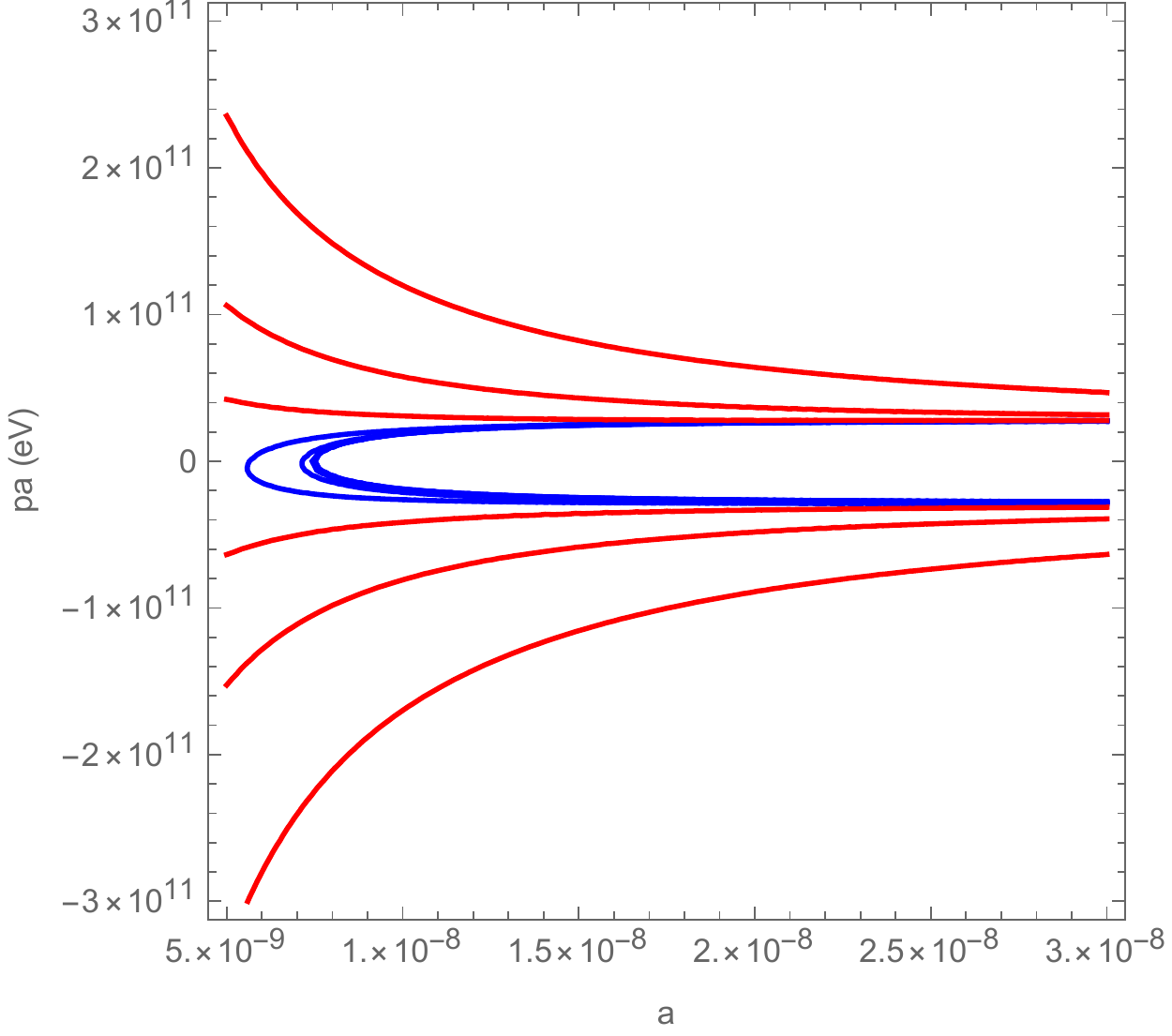}}
	\subfloat{\includegraphics[width=0.5\textwidth]{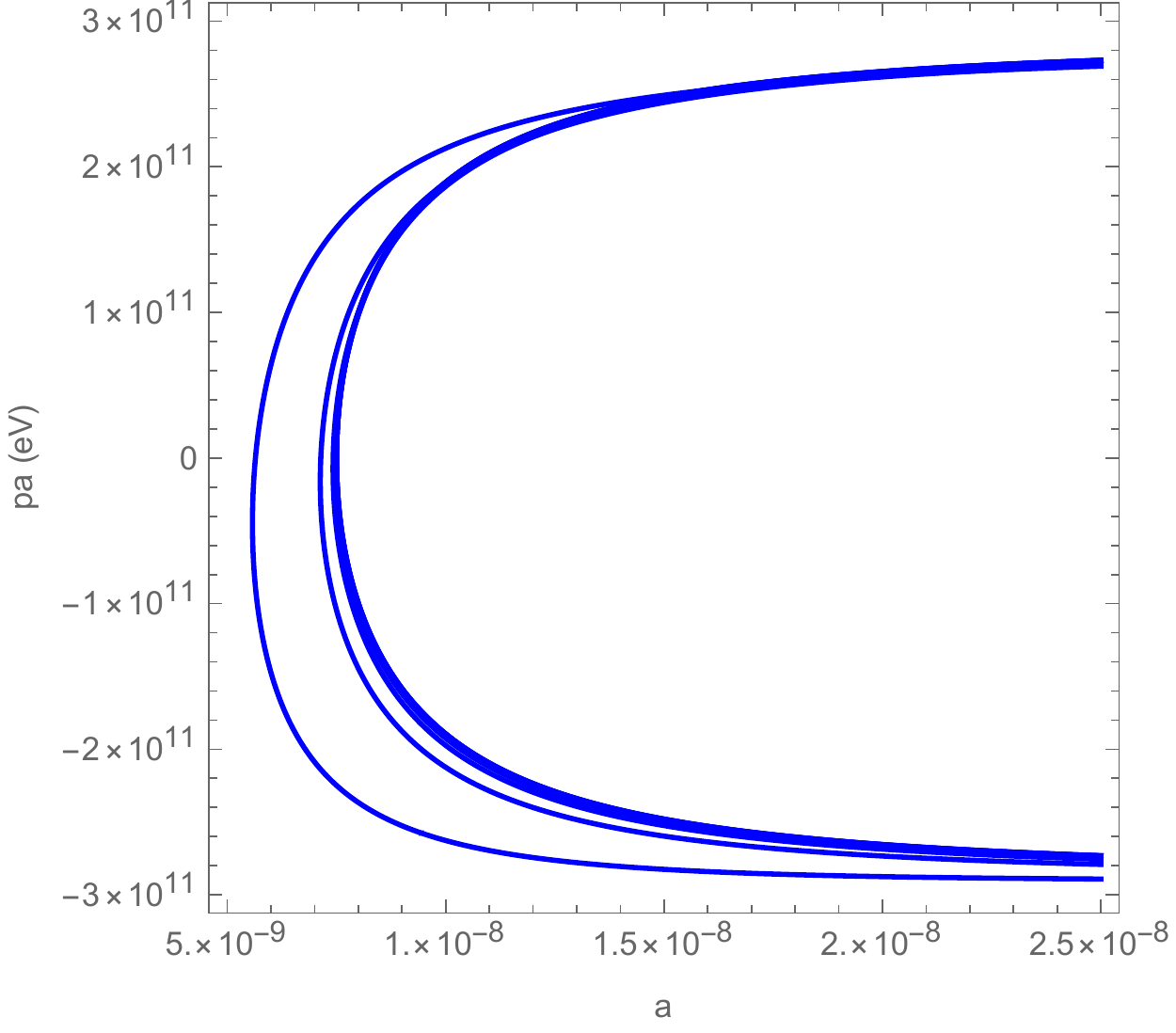}}\hfill
	\caption{The effect of the Brans-Dicke constant in the scalar field phase-space. Once again, we use $E_{0}=10^{16}$ and consider the range $1  \leq p_{\varphi} \leq 10^3$. The left figure is for $\omega=41,000$, and the right figure is for $\omega=4,100,000$.}
	\label{Jordan omega varying}
\end{figure*}

\bigskip
\textbf{Figure \ref{Jordan omega varying}:}	

Now, we study the effect of the Brans-Dicke parameter $\omega$. In the left figure, we take $\omega=41,000$ and see there are more divergent lines than in the generic case considered in Figure \ref{Jordan moment. varying}. In the right figure, we increased $\omega$ to $4,100,000$. Notice that it requires a much greater initial momentum for the scale factor to obtain divergent solutions. Therefore, a larger $\omega$ seems to lead to a more well-behaved theory. 
This is a result of interest, since the larger $\omega$ is, the greater the coupling between matter and the scalar field, that is, the smaller the effects of the scalar field are. This would correspond to the weak-field limit we observe today. Actually, for a perfect fluid (as in our case), we recover GR in this limit \cite{Paiva}. 

\begin{figure*}[htp!]
	\centering
	\figuretitle{Phase-space portrait for $p_{a}$ and $a$, varying the value of $E$.}
	\subfloat{\includegraphics[width=0.5\textwidth]{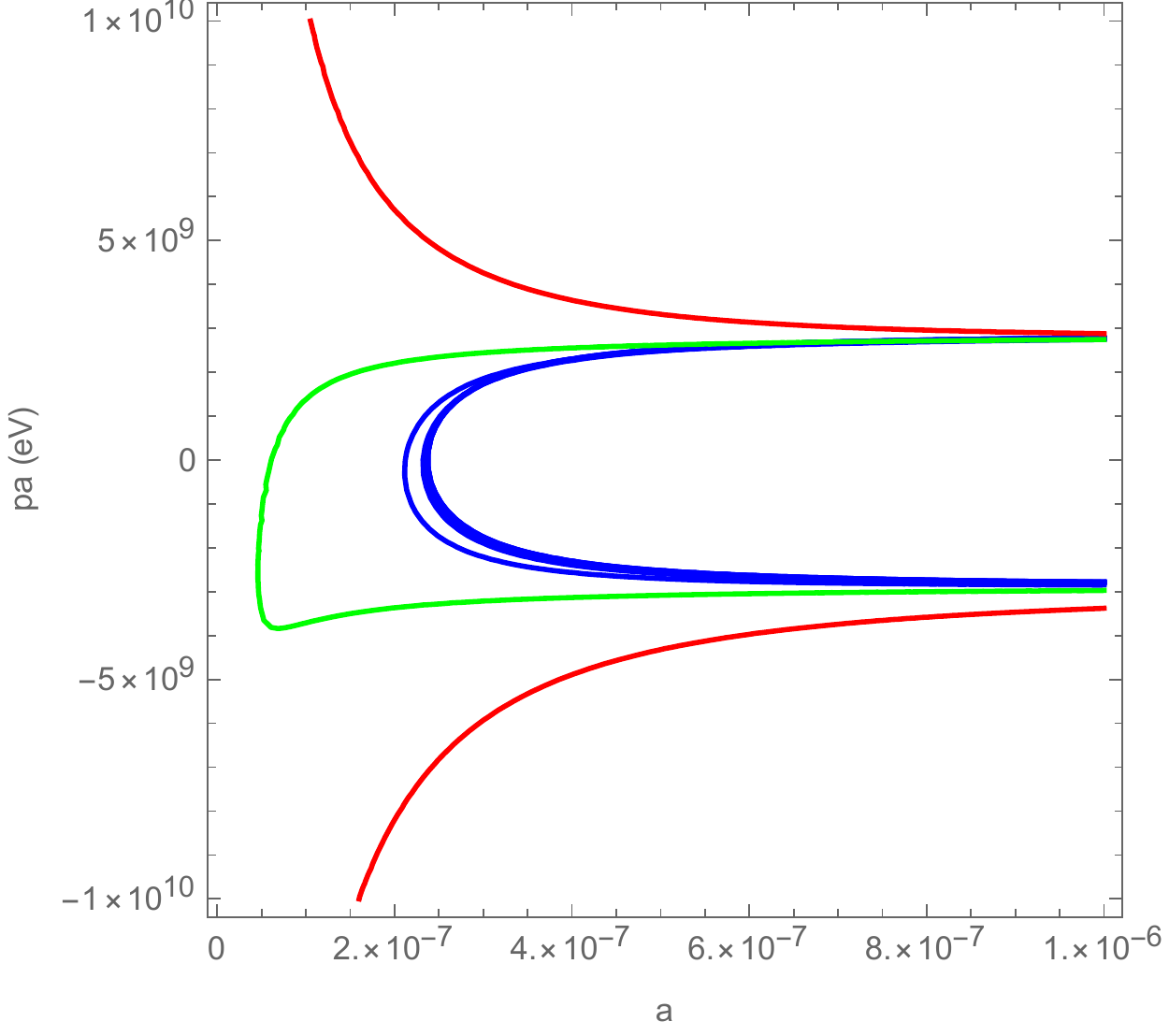}}
	\subfloat{\includegraphics[width=0.5\textwidth]{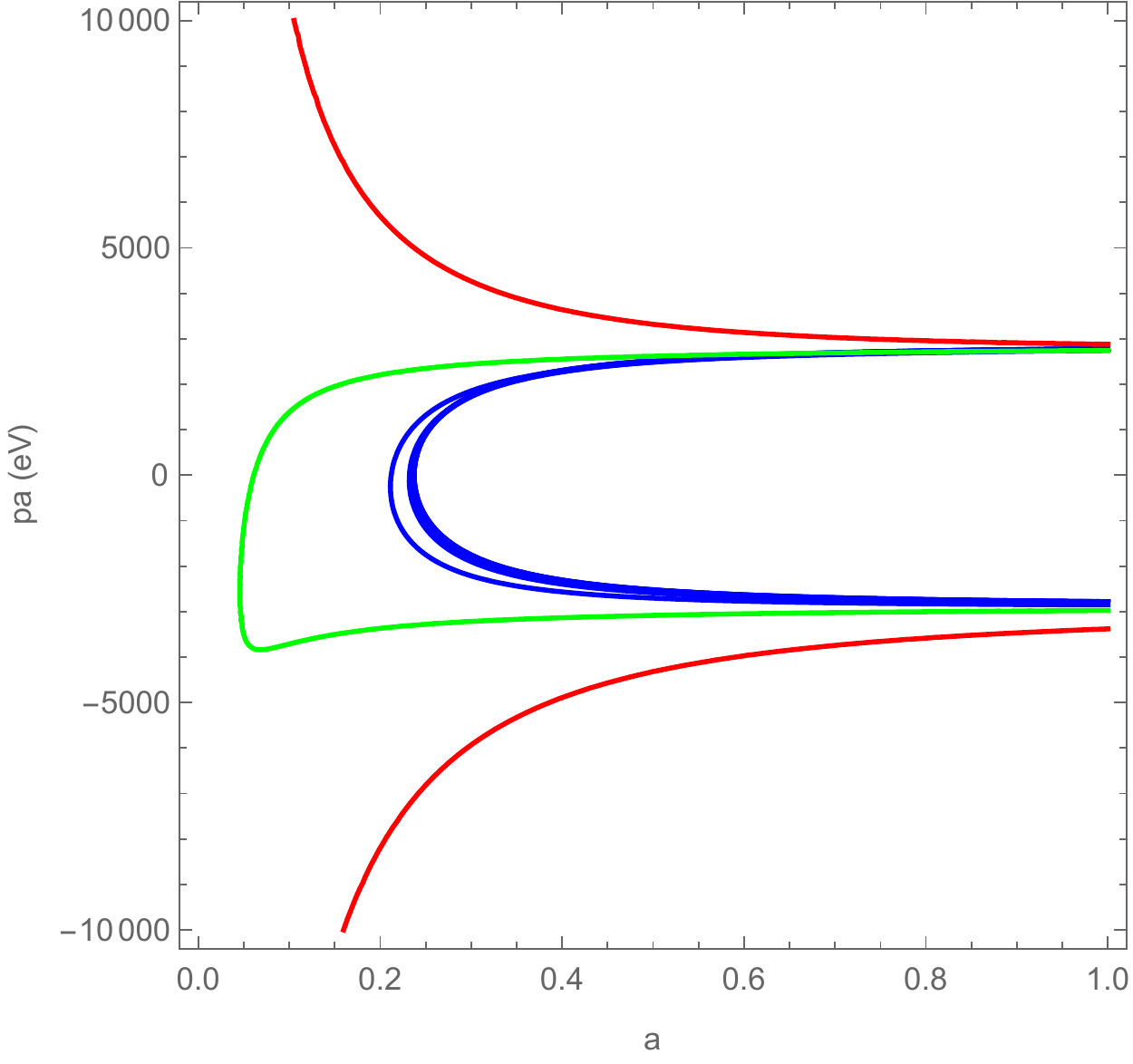}}\hfill
	\caption{The change in the energy of the system results in a change of scale for the solutions. In the left figure we take $E=10^{13}$ and in the right figure we take $E=10$. The same values were used as before for $p_{\varphi}$ and $\omega$: $1  \leq p_{\varphi} \leq 10^3$ and $\omega=410,000$.}
	\label{Jordan E varying}
\end{figure*}

\bigskip
\textbf{Figure \ref{Jordan E varying}:}

The variation of the energy parameter does not change the behavior of the solutions, but it results in a change of scale in the phase-space. So the energy can determine the scale with which inflation happens.

\begin{figure*}[htp!]
	\centering
	\figuretitle{Phase-space portrait for $p_{a}$ and $a$, varying the initial value of the scalar field $\varphi_{0}$.}
	\subfloat{\includegraphics[width=0.5\textwidth]{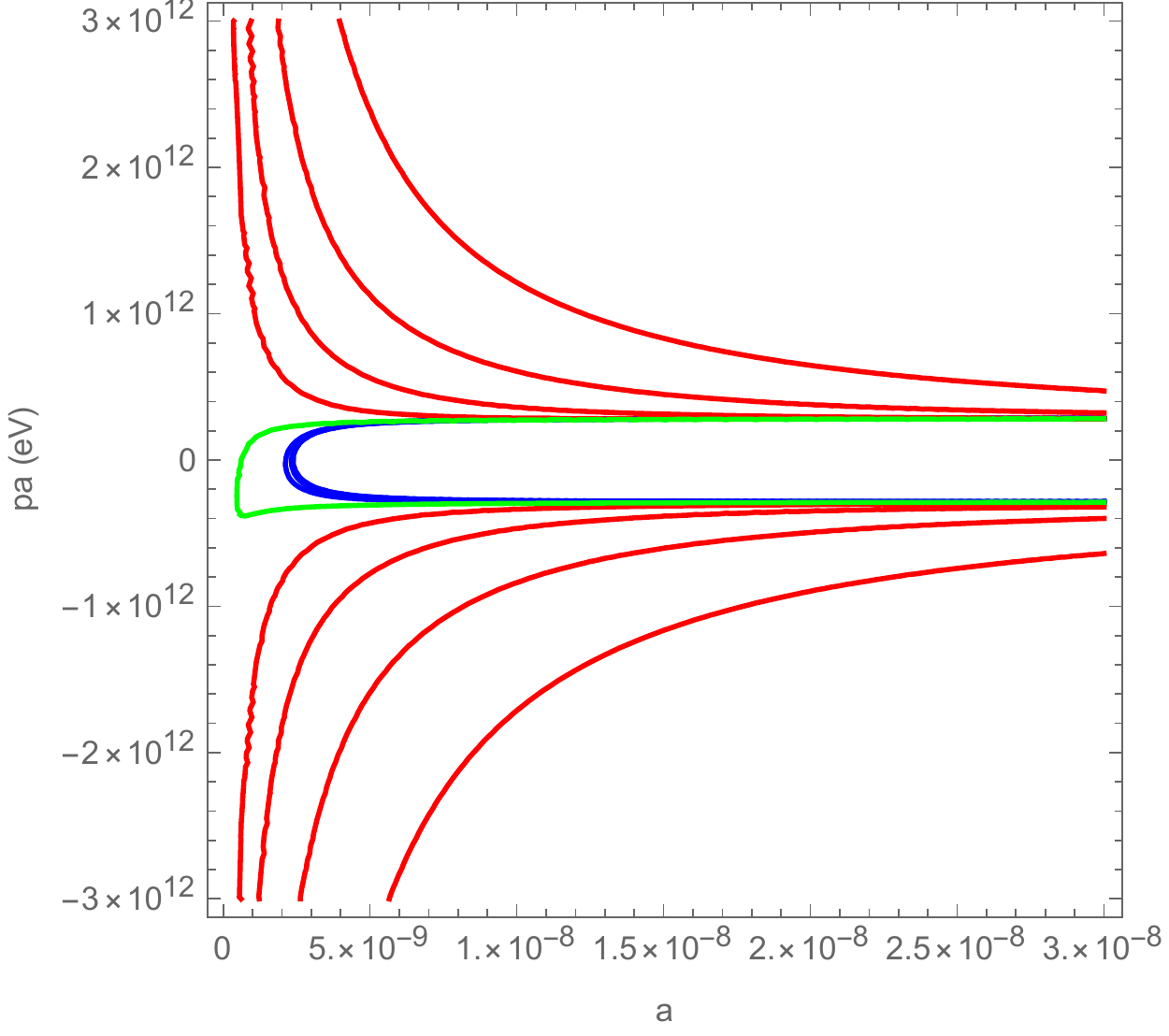}}
	\subfloat{\includegraphics[width=0.5\textwidth]{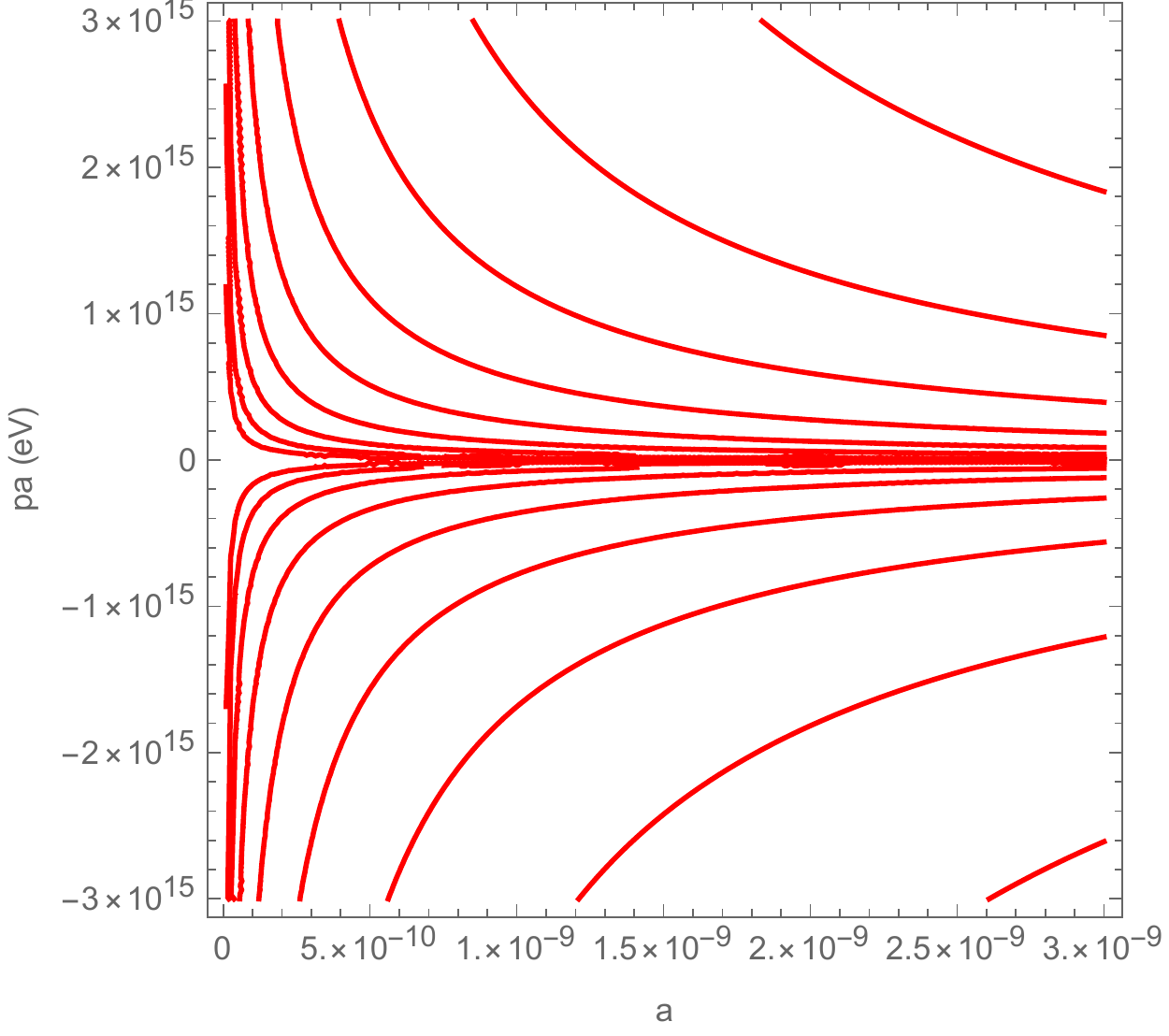}}\hfill
	\subfloat{\includegraphics[width=0.5\textwidth]{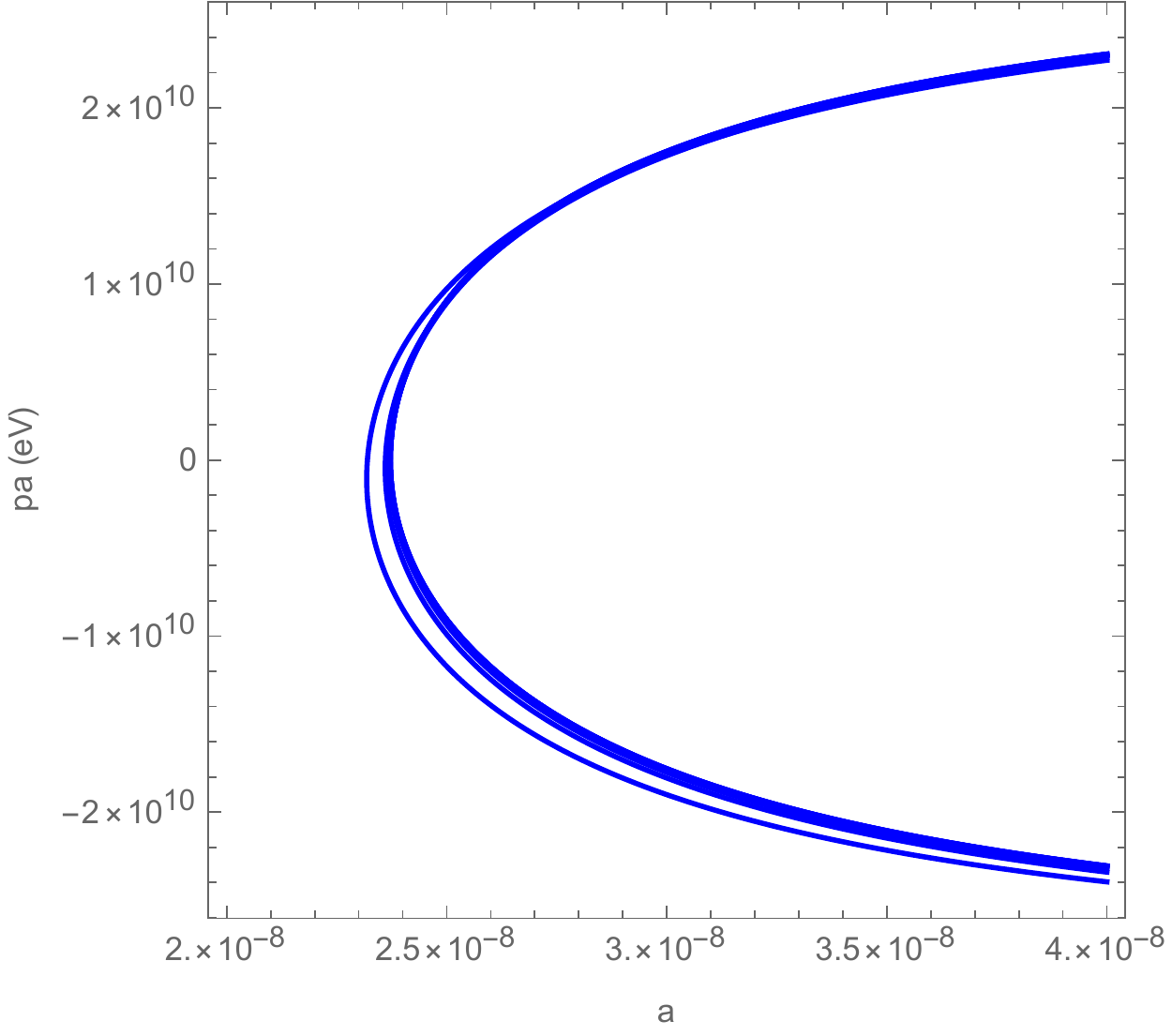}}
	\subfloat{\includegraphics[width=0.5\textwidth]{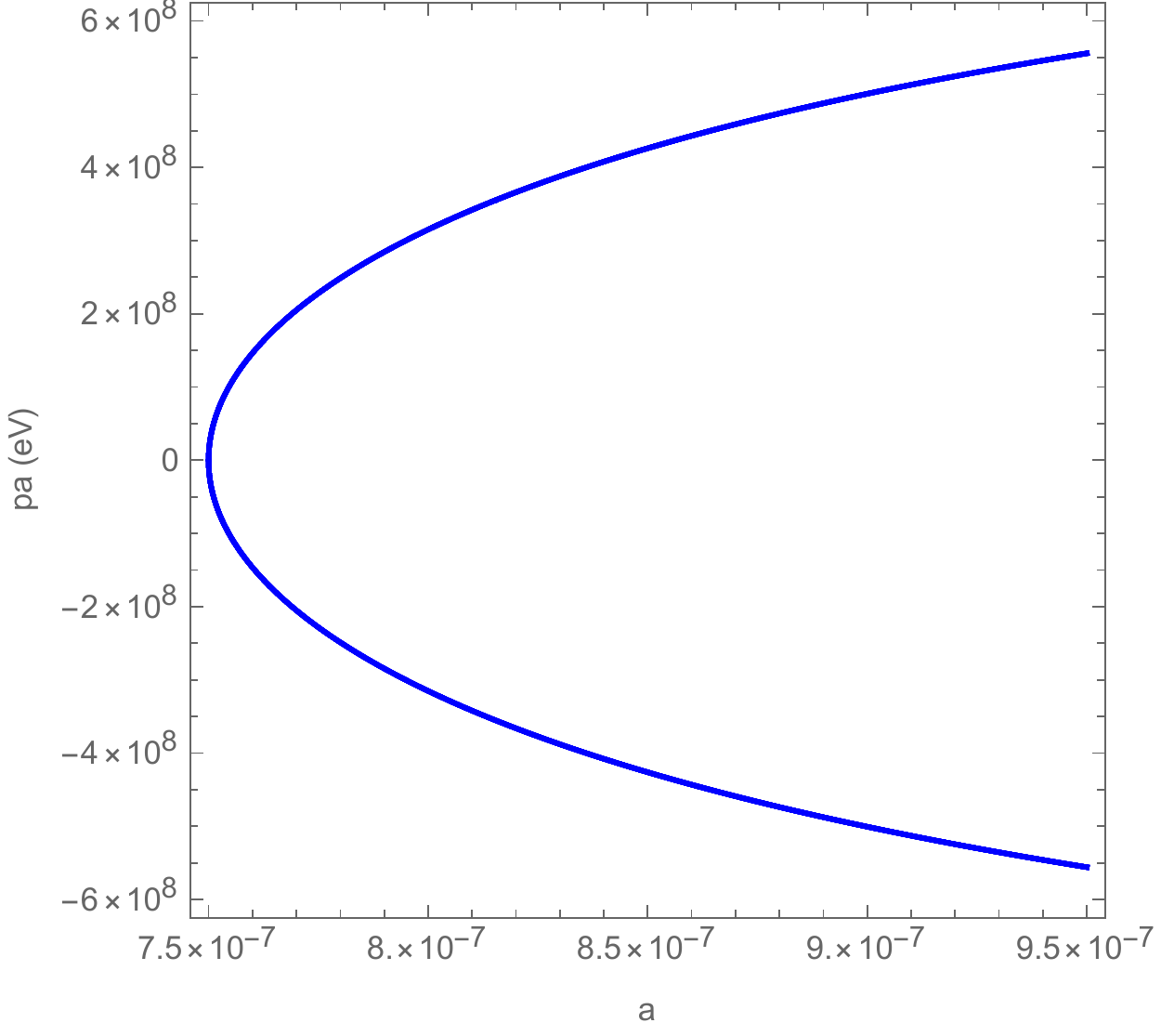}}\hfill
	\caption{The top row shows the solutions for high values of $\varphi_0$: top-left $\varphi_0 = 10$, and top-right $\varphi_0 = 10^4$. The bottom row is for low values of $\varphi_0$: bottom-left $\varphi_0 = 10^{-1}$, and bottom-right $\varphi_0 = 10^{-4}$. For these, we are considering $\omega=410,000$, $E=10^{16}$, and $1  \leq p_{\varphi} \leq 10^3$.}
	\label{plot_pa_phi}
\end{figure*}

\bigskip
\textbf{Figure \ref{plot_pa_phi}:}

Up until now, we have considered the initial value of the scalar field to be $\varphi_{0} =1$, but we also want to understand the effects of the initial condition on the behaviour of the solutions. Thus, in Figure \ref{plot_pa_phi}, we show the direct influence of changing the value for the scalar field on the solutions. The top row shows greater values for $\varphi_0 $, from $10$ to $10^4$ (left to right). We notice that the greater $\varphi_0$ is, the more singularities we obtain. Conversely, in the second row, we lower it from $0.1$ to $10^{-4}$. The solutions tend to bounces instead of singularities. As expected, the results are consistent with the study on $\omega$. 

\subsection{Einstein frame}

\begin{figure*}[htp!]
	\centering
	\figuretitle{Phase-space portrait for $p_{b}$ and $b$, varying the range of $p_{\varphi}$.}
	\subfloat{\includegraphics[width=0.5\textwidth]{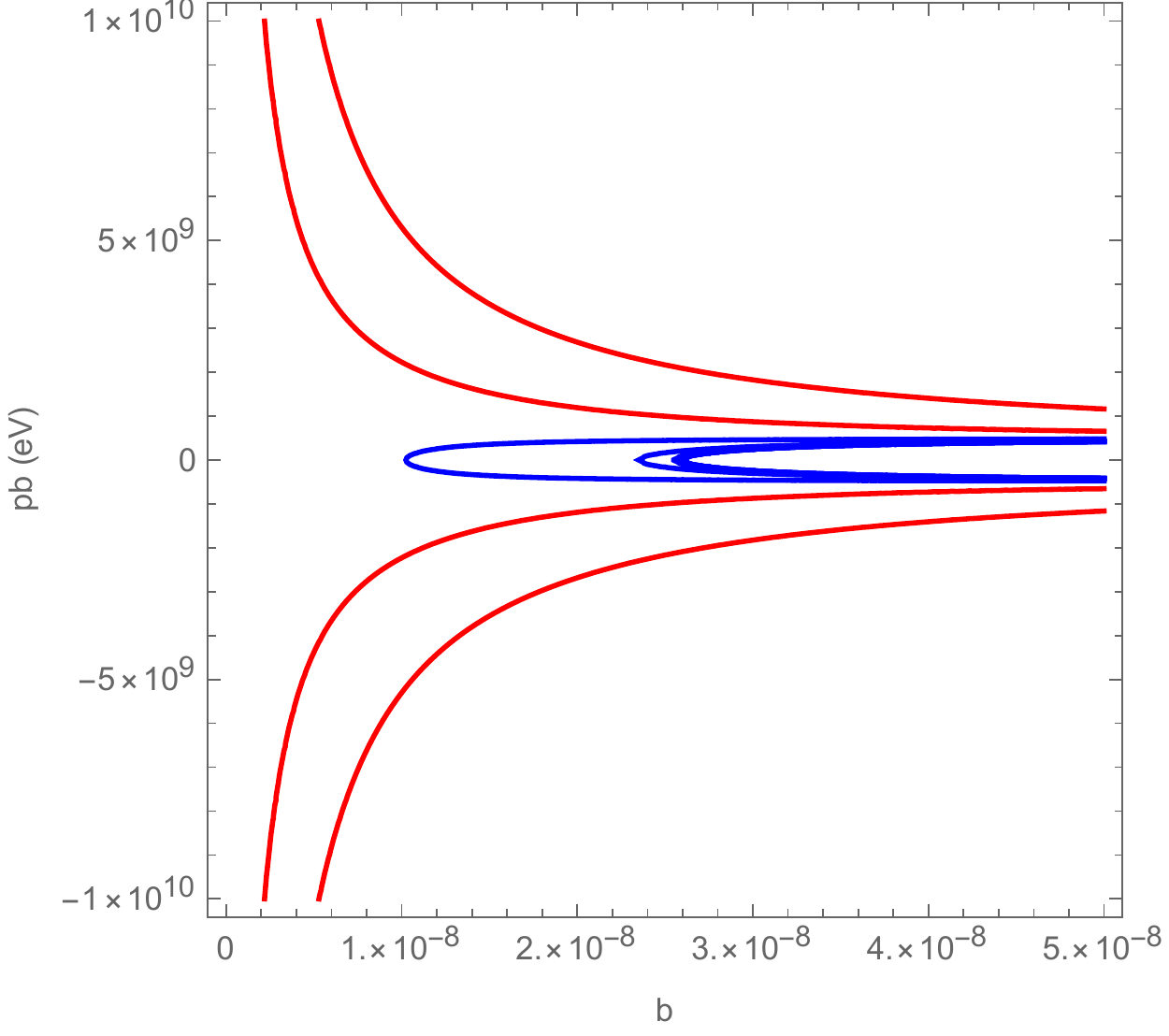}}
	\subfloat{\includegraphics[width=0.5\textwidth]{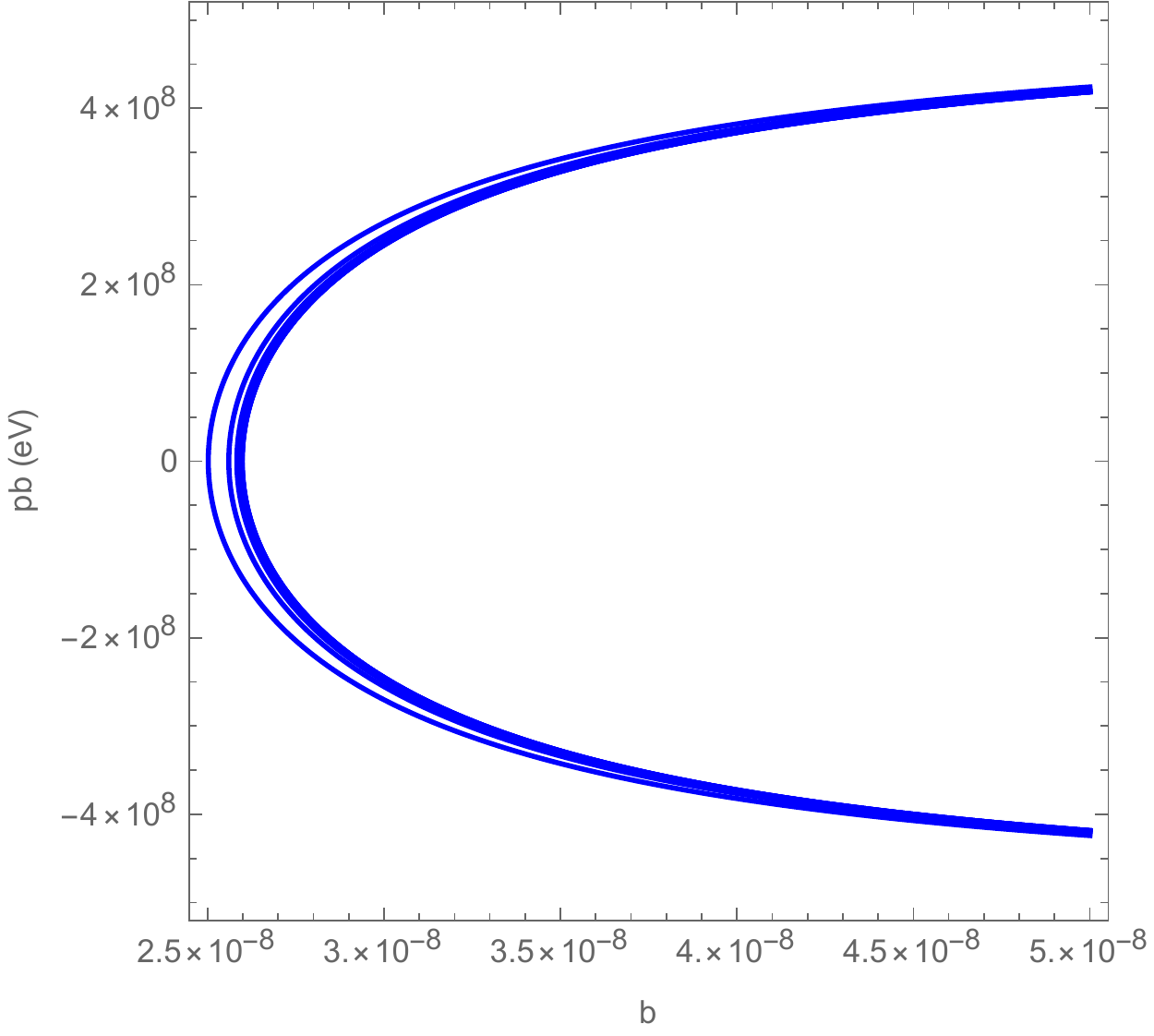}}\hfill
	\caption{Quantum phase-space of the scalar field in the Einstein frame, using $\omega=410,000$ and $E_{0}=10^{16}$. The left figure is for a range $1  \leq p_{\varphi} \leq 10^3$, while for the right figure the  range is smaller $1 \leq p_{\varphi} \leq 10^2$.}
	\label{Einstein moment. varying}
\end{figure*}

\begin{figure*}[htp!]
	\centering
	\figuretitle{Phase-space portrait for $p_{b}$ and $b$, varying the value of $\omega$.}
	\subfloat{\includegraphics[width=0.5\textwidth]{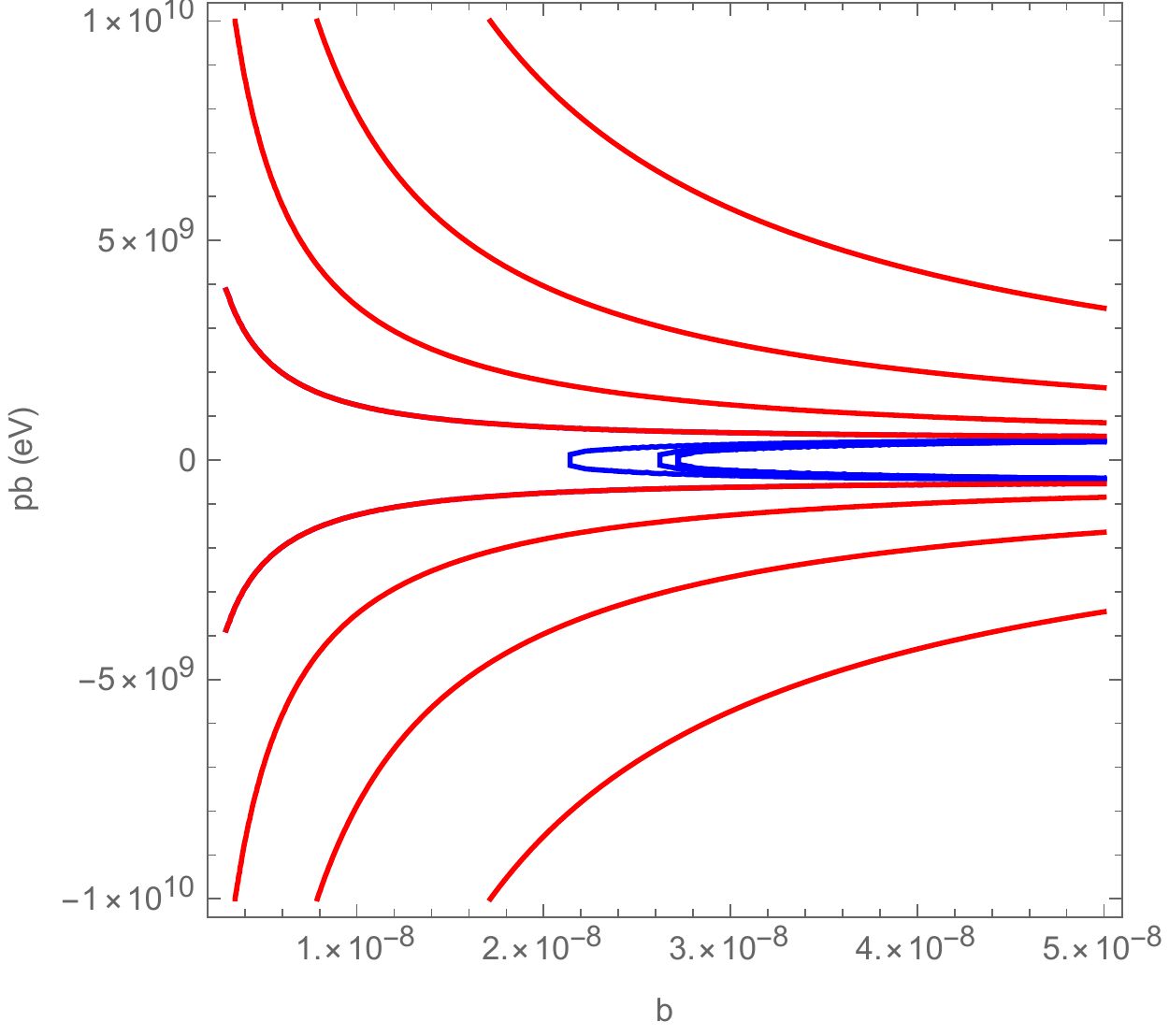}}
	\subfloat{\includegraphics[width=0.5\textwidth]{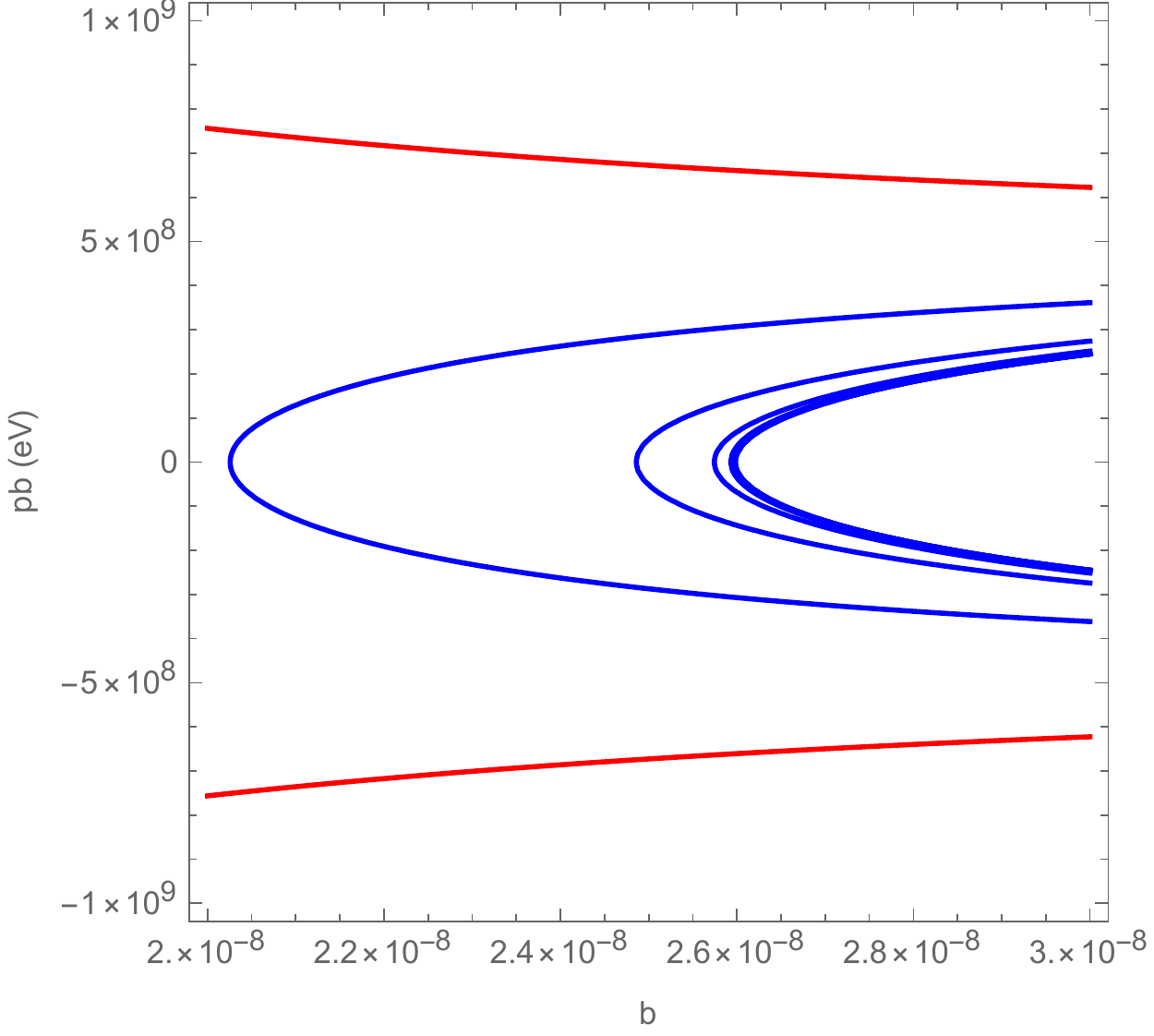}}\hfill
	\caption{The effect of the Brans-Dicke constant in the scalar field phase-space. Once again, we take $E_{0}=10^{16}$ and consider the range $1  \leq p_{\varphi} \leq 10^3$. The left figure is for $\omega=41,000$, and the right one is for $\omega=4,100,000$.}
	\label{Einstein omega varying}
\end{figure*}

\begin{figure*}[htp!]
	\centering
	\figuretitle{Phase-space portrait for $p_{b}$ and $b$, varying the value of $E$.}
	\subfloat{\includegraphics[width=0.5\textwidth]{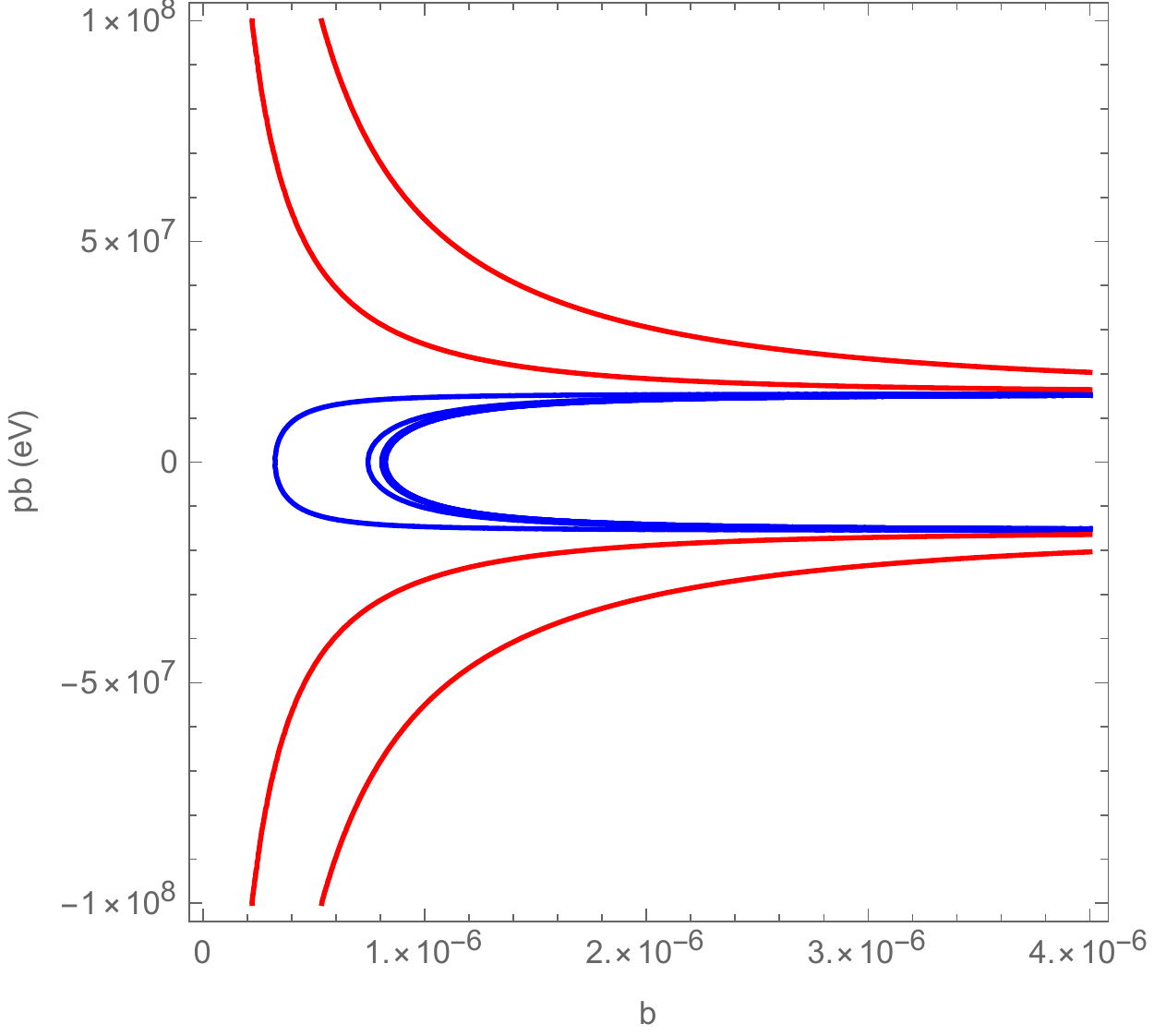}}
	\subfloat{\includegraphics[width=0.47\textwidth]{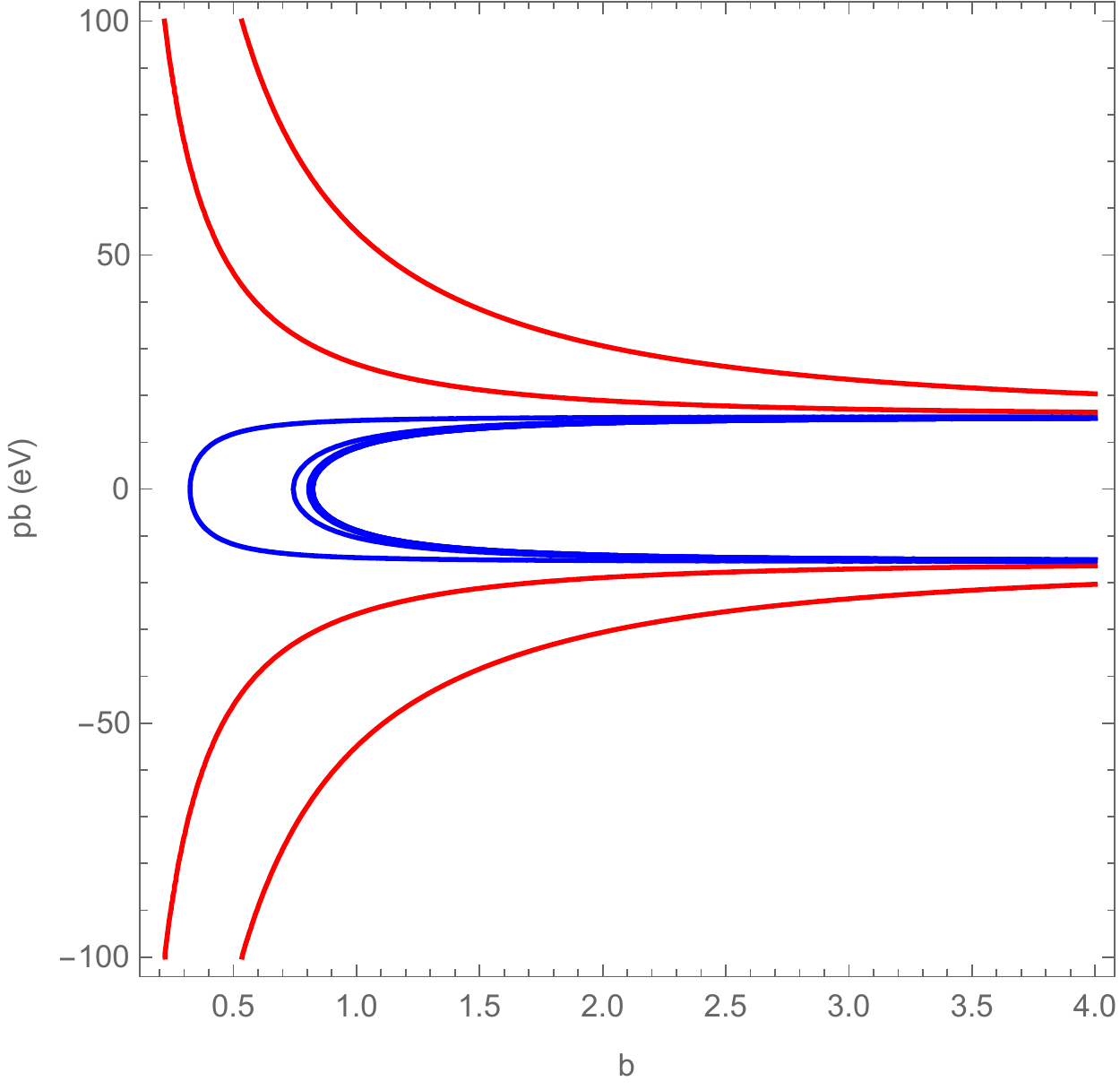}}\hfill
	\caption{The change in the energy of the system results in a change of scale for the solutions. In the left figure we take $E=10^{13}$ and in the right figure we take $E=10$. The same values were used as before for $p_{\varphi}$ and $\omega$: $1  \leq p_{\varphi} \leq 10^3$ and $\omega=410,000$.}
	\label{Einstein E varying}
\end{figure*}

\begin{figure*}[htp!]
	\centering
	\figuretitle{Phase-space portrait for $p_{b}$ and $b$, varying the initial value of the scalar field $\varphi_{0}$.}
	\subfloat{\includegraphics[width=0.5\textwidth]{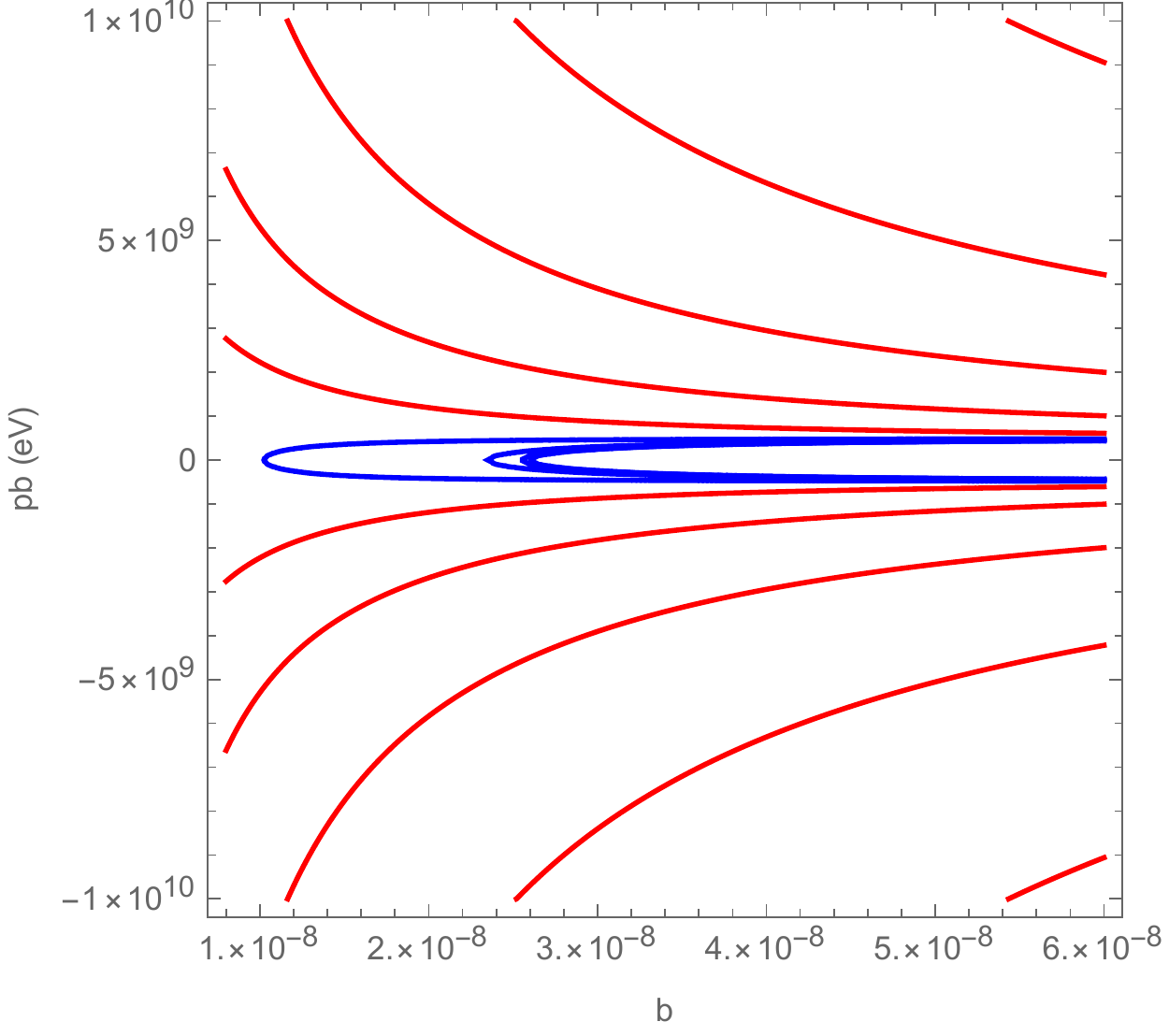}}
	\subfloat{\includegraphics[width=0.5\textwidth]{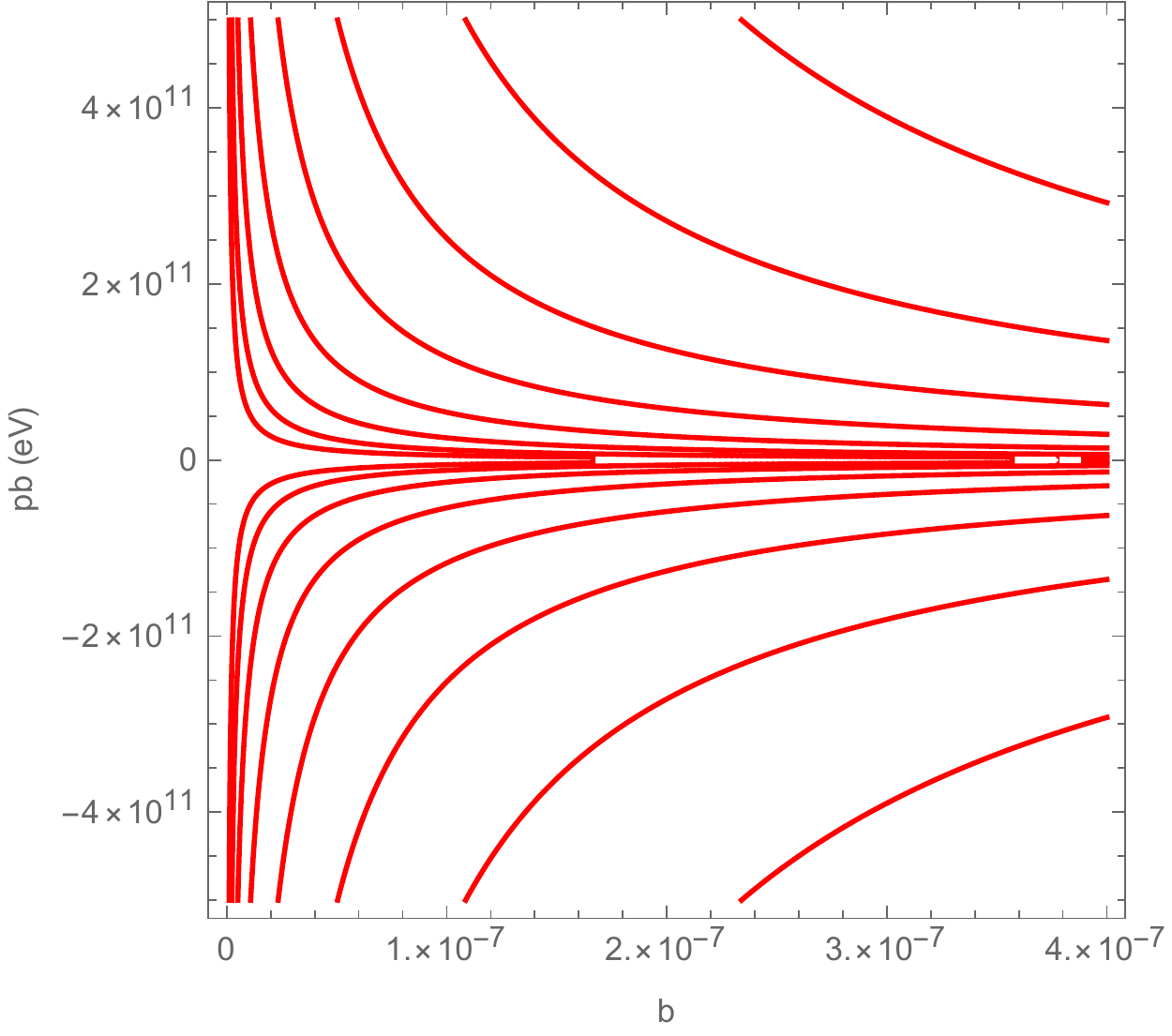}}\hfill
	\subfloat{\includegraphics[width=0.5\textwidth]{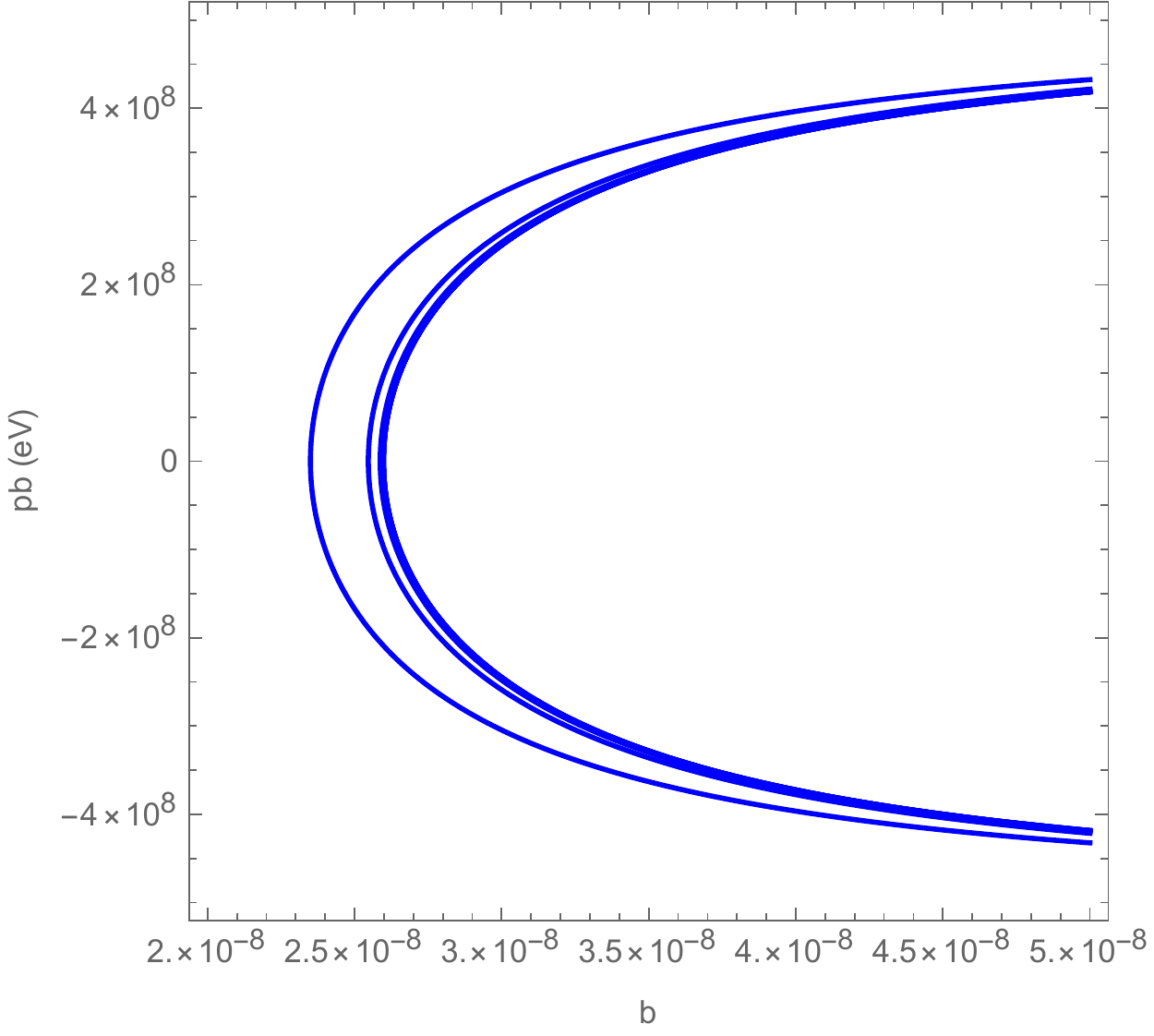}}
	\subfloat{\includegraphics[width=0.5\textwidth]{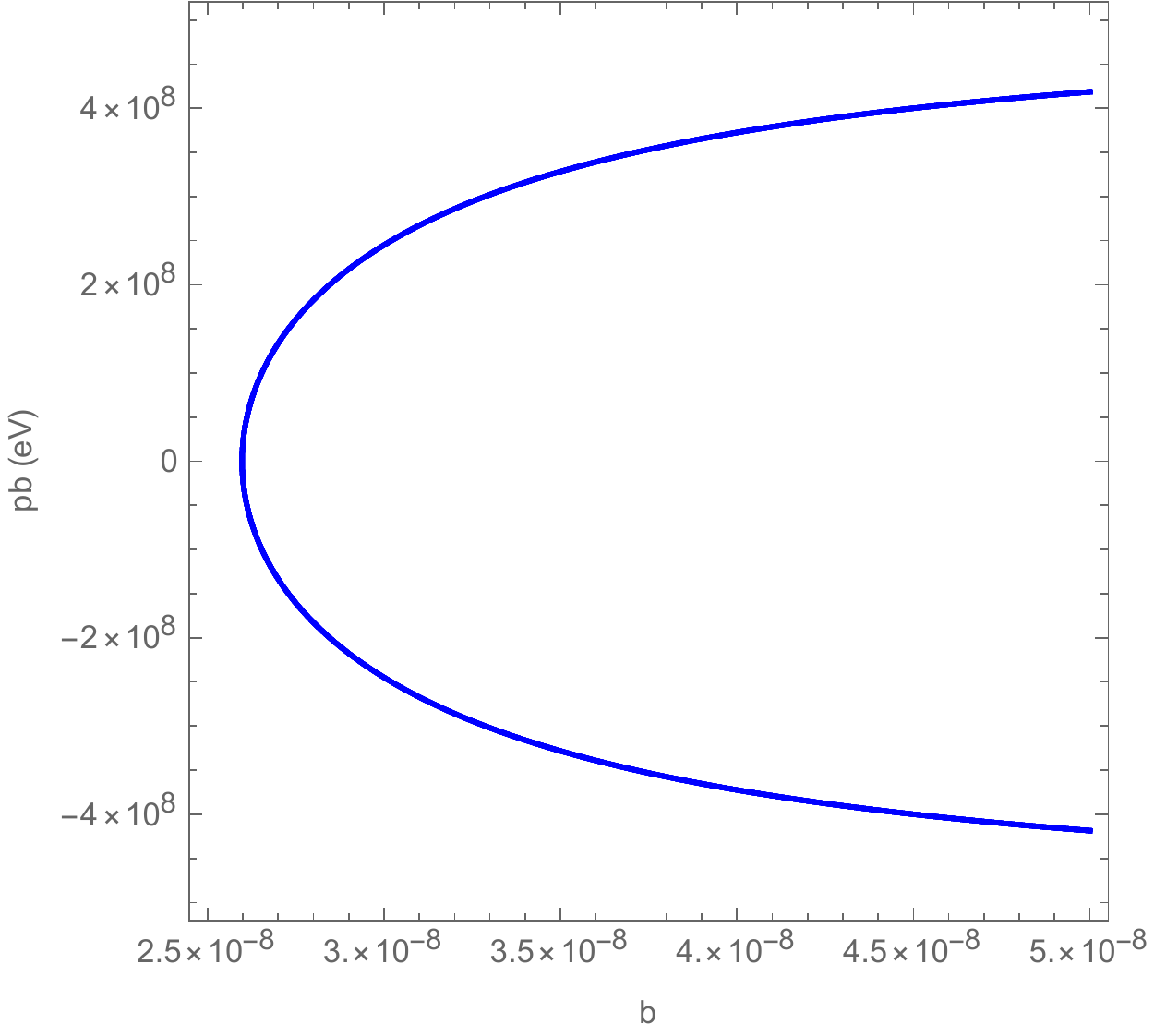}}\hfill
	\caption{The top row shows the solutions for high values of $\varphi_0$: top-left $\varphi_0 = 10$, and top-right $\varphi_0 = 10^4$. The bottom row is for low values of $\varphi$: bottom-left: $\varphi_0 = 10^{-1}$, and bottom-right $\varphi_0 = 10^{-4}$. For these, we are considering $\omega=410,000$, $E=10^{16}$, and $1  \leq p_{\varphi} \leq 10^3$.}
	\label{plot_pb_phi}
\end{figure*}

In the Einstein frame, we have symmetric bounces without any inflationary epoch\footnote{Inflation may be interpreted as a ``stretching" of the solutions induced by the conformal transformation by going from the Einstein frame to the Jordan frame.}, as we see in Figure \ref{Einstein moment. varying}. By varying once again $\omega$ (Figure \ref{Einstein omega varying}) and the energy (Figure \ref{Einstein E varying}), we arrive to the same conclusions as in the Jordan frame, \textit{i.e.} that the larger $\omega$ is, the less divergent the curves we obtain, and varying the energy induces a scaling in the phase space. We also show the effect of the scalar field in Figure \ref{plot_pb_phi}.

Notice that these results are consistent with what was found in the Jordan frame, which provides further evidence that the frames are equivalent. Remember, though, that in spite of choosing specific fiducial vectors, this analysis is still qualitative, since one can always choose different wavelets and also restore the unities (we chose $c=\hbar=1$). For our purpose, this qualitative analysis is enough.

\section{Conclusions} 

In this work, we presented the quantisation of the Brans-Dicke Theory using the affine covariant integral method, and the cosmological scenarios arising from it.
We introduced the classical Hamiltonian formalism of the BDT and the mathematical foundations of this quantisation method, in order to familiarise the reader with the concepts used later on. Our model is completed with the addition of a radiative matter component in form of a perfect fluid, introduced via the Schutz formalism, which we adopted as the clock. The affine quantisation is based on the symmetry of the phase-space of the system, and we can choose the free parameters, namely the fiducial vectors, in a way to build an essentially self-adjoint Hamiltonian operator. The quantisation of the Hamiltonian constraint results in the Wheeler-DeWitt equation, from which we obtain a Schr\"odinger-like equation \eqref{WDW equation}, with the radiative matter providing the time parameter. One expected setback of this quantisation is that it results in a non-separable partial differential equation. We can work around this problem by changing frames, making a conformal transformation of the coordinates.

The BDT is described in the Jordan frame and a conformal change of coordinates transforms the BDT into GR with a scalar-field, \textit{i.e.} the Einstein frame. The equivalence between these frames is still debatable (see \textit{e.g.} \cite{Artymowski,Banerjee,Carla4,Kamenschchik,Pandey}), and our results may contribute to this debate. In the Einstein frame, the Schr\"odinger-like equation is separable, and becomes easier to deal with. We presented the classical GR with a scalar-field model corresponding to the BDT in the Einstein frame and quantised it using the affine method. We also performed a change of coordinates in the already quantised Schr\"odinger-like equation in the Jordan frame. Considering the freedom in the choice of the fiducial vectors, we found an equivalent equation. However, we conclude that the Hamiltonian operator in the Einstein frame is only essentially self-adjoint if we consider different fiducial vectors while quantising the theory in each frame, or if we change the domains (i.e. the measure) of the operators in the respective Hilbert space. In any case, one may argue that, because of this, there is no equivalence between the frames. However, the role of the fiducial vectors during the quantisation is precisely to open up opportunities for adjustment, since it is based on a statistical method ($|\langle q,p| \phi \rangle|^{2}$ is interpreted as the probability density distribution of the funtion $\phi$, see for example \cite{Gazeau1}). Thus, considering different fiducial vectors in different frames should not invalidate the equivalence between them. We choose to solve the Wheeler-deWitt equation obtained from the classical BDT in the Einstein frame, in order to do a qualitative analysis, since this equation has a relatively simple solution. From it, we were able to conclude that the energy spectrum of the Hamiltonian operator in Einstein frame is discrete.

The affine quantisation method is completed with a ``de-quantisation", known as the quantum phase-space portrait or lower symbol, that transforms the quantised operator into a classical function, by means of their fiducial vectors expectation values. This de-quantisation provides a quantum correction for classical observables, from which we can analyse the behavior of these observables in semi-classical environments. Even if we cannot find the wave-function of the Universe in the Jordan frame, we can use the  quantum phase-space to compare the results with the ones from the quantum phase-space in Einstein frame. Thus, we find quantum corrections for the Hamiltonian constraint in both frames in section \ref{Quantum phase-space BDT} and compare the results in Section \ref{Results}, drawing the phase-space portrait for the scale factor, to better understand the behaviour of the (volume of the) Universe in earlier stages.

We obtained two types of solutions in both frames: bounces and singularities. For both types, we predict a prior universe. For the singular cases in the Jordan frame, there is an accelerated contraction, with a singular point where the volume of the Universe becomes null, followed by a decelerated inflationary era. However, if we limit the momentum of the scalar field, we obtain only bouncing solutions. Thus, we may argue that the scalar field should have a limited velocity, since this discards the singular solutions. We also analysed the influence of other parameters in the solutions. In the limit $\omega \rightarrow \infty$, in which we expect to reproduce GR (for our model, at least), bounces become more expected. It is interesting to see that an inflationary stage also appears for bounces in this frame. In the Einstein frame, however, we do not have any inflationary era, but similar conclusions can be drawn, with the exception that both singular and bouncing solutions are symmetric. 

The use of affine quantisation in cosmology is a nascent subject, with the desirable feature of providing solutions without singularities for a natural range of parameters. This is in adequation with other results on various cosmological scenarii (see \cite{Bergeron1,Bergeron2,Bergeron3,Bergeron4,Bergeron7}) as well as a  result on the quantum Belinski-Khalatnikov-Lifshitz scenario using the affine coherent states quantisation for a Bianchi IX universe \cite{Gozdz}, where they suggest that quantising GR should also lead to bouncing solutions. We intend to continue to explore this line of research in future works.

\section*{Acknowledgments}

EF was financed in part by the Coordena\c{c}\~ao de Aperfei\c{c}oamento de Pessoal de N\'ivel Superior - Brasil (CAPES) - Finance Code 001, and by the Institute of Cosmology and Gravitation. EF and CRA thank immensely Prof. Jean-Pierre Gazeau for his support, the referee for his useful comments, and also Chris Pattison for his diligent proofreading of this work.

\end{document}